\documentclass[preprintnumbers,amsmath,amssymb]{revtex4}
\usepackage{epsfig}
\def\w{{\rm w}}
\let\jnfont=\rm
\def\NPB#1,{{\jnfont Nucl.\ Phys.\ B }{\bf #1},}
\def\PLB#1,{{\jnfont Phys.\ Lett.\ B }{\bf #1},}
\def\EPJC#1,{{\jnfont Euro.\ Phys.\ J.\ C }{\bf #1},}
\def\PRD#1,{{\jnfont Phys.\ Rev.\ D }{\bf #1},}
\def\PRL#1,{{\jnfont Phys.\ Rev.\ Lett.\ }{\bf #1},}
\def\MPLA#1,{{\jnfont Mod.\ Phys.\ Lett.\ A }{\bf #1},}
\def\JPG#1,{{\jnfont J.\ Phys.\ G}{\bf #1},}
\def\CTP#1,{{\jnfont Commun.\ Theor.\ Phys.\ }{\bf #1},}
\def\JHEP#1,{{\jnfont J. High \ Ener.\  Phys.}{bf #1},}
\def\RMP#1,{{\jnfont  Rev. Mod. Phys.}{bf #1},}

\def\met{\displaystyle{\not}E_T}

\def\ddbar{d\bar d}
\def\ccbar{c\bar c}
\def\ssbar{s\bar s}
\def\bbbar{b\bar b}

\begin{document}
\preprint{}

\title{Charged Higgs Production in Association With $W^{\pm}$
at Large Hadron Colliders}

\author{Guo-Li Liu$^{1,3}$, Fei Wang$^{1,3}$, Shuo Yang$^{2,3}$}
\affiliation{ $^1$ Physics Department, Zhengzhou University, Henan,
450001, China \\
$^2$ Physics Department, Dalian University, Dalian, 116622, China \\
$^3$ Kavli Institute for Theoretical Physics China, Academia Sinica,
Beijing 100190, China }

\begin{abstract}
Many new physics models beyond the standard model, such as the littlest higgs models and the left right twin higgs models,
predict the existence of the large charged higgs couplings $H^-q\bar b$ and $H^+b\bar q$, where $q=t$ or the new vector-like heavy quark $T$;  On the other hand, some new physics models like the littlest higgs also predict the gauge-higgs couplings. Such couplings may have rich collider phenomenology. We focus our attention on these couplings induced by the littlest higgs models and the left right twin higgs models models and consider their contributions to the production cross section for $W^\pm H^\mp$ production at the large hadron colliders. We find that the cross sections, in the littlest higgs models, on the parton level $gg \to W^\pm H^\mp$ and  $q\bar q \to W^\pm H^\mp$ ($q=u,d,s,c,b$)  may reach tens of several dozen femtobarns in reasonable parameters space at the collision energy of $14$ TeV and that the total cross section can even reach a few hundred femtobarns in certain favored space. While in the left right twin higgs models, the production rates are basically one order lower than these in littlest higgs. Therefore, due to the large cross sections of that in the littlest higgs, it may be possible to probe the charged higgs via this process in a large parameter space.

\end{abstract}

\newpage
\maketitle

\section{Introduction}
The primary goal of the Large Hadron Collider (LHC) at CERN
is to verify the electroweak symmetry breaking mechanism
and to discover or rule out the existence of a higgs boson.
Since both the ATLAS and CMS collaborations have discovered a higgs
boson-like particle with mass of around $125$ GeV at a significance
of $5\sigma$ last year, the goal seems to have been reached
\cite{atlas,cms}. Apart from searches for the higgs boson, there is
an ongoing hunting for signals of physics beyond the standard model
(SM) at the LHC, and hopefully these experiments will shed some
light on physics at the TeV scale.

Due to the incompletion, aesthetical and theoretical
problems such as the famous hierarchy problem and triviality problem
of the SM higgs boson, various new physics models beyond
the SM which try to solve in different ways the previous mentioned problems are proposed.
For example, in the Little higgs (LH)\cite{lh-review} models, the higgs bosons
emerges as the pseudo Nambu-Goldstone bosons associate with the spontaneous breaking
of a global symmetry. In order to implement the collective symmetry breaking mechanism,
new particles such as heavy gauge bosons and top-partners
are introduced. Quadratically divergent corrections contributed by such new particles
to higgs boson masses cancel out those induced by the top quark and gauge boson loops at
one-loop level. Thus no much fine tuning is needed in the LH model with cut off scale ${\cal O}({\rm 10 TeV})$.

Another example is the left-right twin higgs (LRTH) \cite{lrth-review} models.
Again, the higgs bosons in nature are pseudo-Goldstone bosons from spontaneously broken global symmetry.
The higgs bosons obtain masses from the gauge and Yukawa interactions which
break the global symmetry. As the left-right symmetry was imposed to the twin higgs mechanism,
the quadratic terms in the higgs potential respect the global symmetry, and the contributions to the higgs
masses cancel out.
The logarithmically divergent terms, however, are radiatively generated which
are not invariant under global symmetry and contribute masses to the pesudo-Goldstones.
 The resulting higgs mass is in the field of the electroweak scale with the cutoff at about $5\sim 10$ TeV.

The LH models and the LRTH models predict multiplet physical higgs bosons, two of which are charged one.
Since it is hard to distinguish between the CP-even higgs bosons
in such new physics models and the higgs boson in the SM, any observation of a
charged higgs will be a crucial signature for new physics beyond the SM.
That is why the charged scalar particles have attracted much
attention in the previous years by different high energy physics
experiments and theories and they will certainly be probed at the LHC.

The search for higgs bosons and new physics particles and the study
of their properties are among the prime objectives of the large
hadron collider (LHC) \cite{Quigg2007}.
Since the discovery of the charged higgs bosons will be the evidence of
new physics beyond the SM, there are increasing interests in
theoretical and experimental studies to provide the basis for its
accurate exploration. Therefore the LH and LRTH models are very interesting since in these
models charged scalars are predicted and they may
possess larger tree-level or one-loop top or bottom Yukawa couplings
so we may detect the new Yukawa coupling in these models, which may
serve as a sensitive probe of the two models.

Much effort is put in the search for charged higgs bosons.
From its exclusive decay modes  $H^{\pm}\to
\tau\nu$ and $H^{\pm}\to cs$, the LEP search experiments   \cite{LEP,higgs-lep}
have given a direct detection limit $M_{H^{\pm}}>78.6$ GeV.
with different mass range, the search approaches for charged higgs bosons at the
hadron collider are distinct.
When the charged higgs mass is low, the signal will be $t\to H^+b\to \bar{\tau}\nu_\tau b$.
The Tevatron search is mainly focused on the low mass range
$m_{H^{\pm}}<m_t$ due to phase space suppression for a heavy charged higgs boson
production and any signal of the charged higgs has not been found  which means that
the mass of the  charged higgs is larger than $160$ GeV \cite{Tevatron}.
At the LHC, however, the charged higgs search  can be
feasible via such as the $gb\to tH^-$ and $gg \to t\bar bH^-$ production
up to a large mass
range since the LHC collision energy is large \cite{TDR}.
When the charged higgs is heavy $m_{H^{\pm}}>m_t$, the signal is from the main production
process $gb\to tH^-$ and $gg \to t\bar bH^-$ followed by its main decay
$H^-\to t\bar{b}$  \cite{bg-h,th-han}.

Motivated by new technique of "jet substructure"
\cite{Jetography,VermilionThesis,boost2010,SpannowskyRev,11114530}
developed for highly boosted massive particles, a "hybrid-R
reconstruction method", which can use the top tagging and the $b$
tagging for other isolated $b$ jets as well as the full
reconstructed objects in the final state to suppress the background,
is proposed to investigate the full hadronical decay channel of the
 heavy charged higgs production.

 Recently, discussions on neutral or charged higgs production
at the LHC have been carried out, see e.g, Refs
\cite{Jakobs2009,1010.4139,1001.0473,0901.3380,1207.4071,1105.2607,1109.2376}.
Both the LH and LRTH models predict neutral or charged
($\phi^0$, $\phi^\pm$ or $H^{\pm}$) scalars with large Yukawa
couplings to the third generation quarks in addition to a SM-like higgs.
They also predict one vector-like heavy top quark $T$
and new gauge bosons ($A_H, Z_H, W_H$). Such new particles can
be regarded as a typical feature of those models. Signals of this two models
 have already been studied in the work environment of linear colliders and hadron-hadron colliders
\cite{lh-lrth-pheno}, but most attentions had been concentrated on the
neutral scalars and new gauge bosons. Here we wish to discuss the
charged scalars aspects.

For the production of charged scalar in association with a $W$
boson at the LHC, there are mainly two kinds of the partonic
subprocesses that contribute to the hadronic cross section $pp
\to W^{\pm}\phi^{\mp}$: the  $q\bar q$ ($q=u,d,c,s,b$)
annihilation and the $gg$ fusion. One has studied processes like these,
in the minimal supersymmetric standard model(MSSM), for example,
many efforts have been made about the $\phi W$ productions \cite{mssm-wh-lhc}
and they may constraint our results.  In this paper, to probe the Littlest
higgs models and the left right twin higgs models,
we shall discuss the production of charged scalar $\phi^{\pm}$
in association with SM
gauge bosons $W^{\mp}$ via those two kinds subprocesses.

This work is organized as follows. In Sec. II we recapitulate the LH
models, give the couplings relevant to our calculation and then
discuss the numerical results in it. Similarly,  in Sec. III the
LRTH models are simply described and the numerical results will be
given.  Finally, we compare the results predicted by the two models
and give the conclusion in Sec. IV.

\section{\bf  The $LH$ model and $W^\pm H^\mp$ production
at the $LHC$ }
\subsection{The $LH$ model and the relative couplings}

The littlest higgs model\cite{littlest} is based on the
$SU(5)/SO(5)$ nonlinear sigma model.
The global symmetry breaks from $SU(5)$ to $SO(5)$, generating 14 Goldstone
bosons, and the gauge symmetry from $[SU(2)\times U(1)]^{2}$  to
$SU(2)\times U(1)$, the SM electroweak gauge group.
Four of these Goldstone bosons are eaten by the broken gauge
generators, resulting in four massive gauge bosons $A_{H}$, $Z_{H}$ and
$W^{\pm}_{H}$.  The left 10 states transform under the SM gauge
group as a doublet H and a triplet $\Phi$. The doublet VEV further breaks
the gauge symmetry $SU(2)\times U(1)$
into $U(1)_Y$, eating three scalar of it and leaving only a CP-even higgs,
usually regarded as the SM higgs, which has been discovered at the LHC \cite{atlas,cms}.

When the fields rotate to the mass eigenstates, the gauge bosons mix by the mixing angle $s$ and $s'$,
\begin{eqnarray}
    s = \frac{g_2}{\sqrt{g_1^2+g_2^2}}, \qquad
    s^{\prime} = \frac{g_2^{\prime}}{\sqrt{g_1^{\prime 2}+g_2^{\prime 2}}}.
\end{eqnarray}

In the LH models \cite{littlest}, a new set of heavy vector-like fermions,
 $\tilde{t}$ and  $\tilde{t^{'c}}$ is introduced in order to
cancel the top quark quadratic divergence, since they couple to higgs field.
Choosing the Yukawa form of the  coupling of the SM top quark to the
pseudo-Goldstone bosons and the heavy vector pair in the LH models, then diagonalizing the mass terms,
one can straightforwardly
work out the higgs-quark interactions, as given in Ref. \cite{littlest}, and we repeatedly list here in Table
\ref{gauge-scalar-fermion} for convenience.

The couplings, related to  our calculation, of the new
particles to the SM particles, which include 1) the three-point
couplings of the gauge boson to the scalars, including case I: one
gauge boson to two scalars and case II: two gauge bosons and one
scalar; 2)charged gauge
boson-fermion couplings; 3)the scalar-fermion couplings, which can be found in Ref. \cite{littlest},
are extracted here as:
\begin{table}[h]
\begin{tabular}{|c|c||c|c|}
\hline
particles & vertices & particles & vertices \\
\hline $W_{L\mu}^+ H \Phi^-$ &  $-\frac{ig}{2} \left( \sqrt{2} s_0 - s_+ \right) (p_1-p_2)_{\mu}$ &
 $W_{L\mu}^+ A_{H\nu} \Phi^-$ & $-\frac{i}{2} g g^{\prime}
    \frac{(c^{\prime 2}-s^{\prime 2})}{2s^{\prime}c^{\prime}}
    ( v s_+ - 4 v^{\prime} ) g_{\mu\nu}$ \\
 \hline $W_{L\mu}^+ \Phi^0 \Phi^-$ & $-\frac{ig}{\sqrt{2}} (p_1-p_2)_{\mu}$  &
 $W_{L\mu}^+ Z_{L\nu} \Phi^-$ & $-i \frac{g^2}{c_\w} v^{\prime} g_{\mu\nu}$ \\
    \hline $W_{L\mu}^+ \Phi^P \Phi^-$ &    $\frac{g}{\sqrt{2}} (p_1-p_2)_{\mu}$ &
 $W_{L\mu}^+ Z_{H\nu} \Phi^-$ & $ i g^2 \frac{(c^2-s^2)}{2sc} v^{\prime} g_{\mu\nu}$  \\
\hline $W_L^{+\mu} \bar t_L b_L$ &
    $\frac{ig}{\sqrt{2}}
    \left[ 1 - \frac{v^2}{f^2} \left( \frac{1}{2} x_L^2
    + \frac{1}{2} c^2 (c^2-s^2) \right) \right]
    \gamma^{\mu} V_{tb}^{\rm SM}P_L$ &
$W_L^{+\mu} \bar T_L b_L$ &
    $\frac{g}{\sqrt{2}}
    \frac{v}{f} x_L
    \gamma^{\mu} V_{tb}^{\rm SM}P_L$ \\
\hline $H \bar t t$ &
    $- i \frac{m_t}{v} \left[ 1 - \frac{1}{2} s_0^2
    + \frac{v}{f} \frac{s_0}{\sqrt{2}} - \frac{2 v^2}{3 f^2} \right.$ &
$H \bar T T$ &
    $- i \frac{\lambda_1^2}{\sqrt{\lambda_1^2 + \lambda_2^2}}
    \left( 1 + \frac{\lambda_1^2}{\lambda_1^2 + \lambda_2^2}\right)
    \frac{v}{f}$ \\
 &
    $\left.
    + \frac{v^2}{f^2} \frac{\lambda_1^2}{\lambda_1^2 + \lambda_2^2}
    \left( 1 + \frac{\lambda_1^2}{\lambda_1^2 + \lambda_2^2} \right)
    \right]$ &
 & \\
\hline
$\Phi^0 \bar t t$ &
    $-\frac{im_t}{\sqrt{2}v}
    \left( \frac{v}{f} - \sqrt{2} s_0 \right)$ &
$\Phi^P \bar t t$ &
    $- \frac{m_t}{\sqrt{2}v}
    \left( \frac{v}{f} - \sqrt{2} s_P \right) \gamma^5$ \\
\hline
$\Phi^+ \bar t b$ &
    $-\frac{i}{\sqrt{2}v}
    \left( m_t P_L + m_b P_R \right) \left( \frac{v}{f} - 2 s_+ \right)$ &
$\Phi^+ \bar T b$ &
    $-\frac{im_t}{\sqrt{2}v} \left(\frac{v}{f} - 2 s_+ \right)
    \frac{\lambda_1}{\lambda_2} P_L$ \\
\hline
\end{tabular}
\caption{the three-point couplings of the gauge boson,  the
scalars, the fermions in the littlest higgs models. The momenta
are taken as $V_{\mu} S_1(p_1) S_2(p_2)$ and the particles
are in the mass eigenstates with the momenta out-going.
\label{gauge-scalar-fermion} }
\end{table}

Here $P_L=(1-\gamma^5)/2$ and $x_L \equiv \lambda_1^2/(\lambda_1^2+\lambda_2^2)$,
where $\lambda_1,~\lambda_2$ are the Yukawa
coupling of order ${\cal O}(1)$.  

The neutral gauge boson-fermion couplings can also be extracted  as those
 in Table \ref{gauge-fermion2} \cite{littlest}.
\begin{table}[hbt]
\begin{tabular}{|c|c|c|}
\hline particles & $g_V$ & $g_A$ \\
\hline $Z_L \bar u u$ &
    $-\frac{g}{2c_{\w}} \left\{ (\frac{1}{2} - \frac{4}{3} s^2_{\w})
    - \frac{v^2}{f^2} \left[ c_{\w} x_Z^{W^{\prime}} c/2s
    \right. \right.$ &
    $-\frac{g}{2c_{\w}} \left\{ -\frac{1}{2}
    - \frac{v^2}{f^2} \left[ -c_{\w} x_Z^{W^{\prime}} c/2s
    \right. \right.$ \\
$$ &
$\left. \left.
    + \frac{s_{\w} x_Z^{B^{\prime}}}{s^{\prime}c^{\prime}}
    \left( 2y_u + \frac{7}{15} - \frac{1}{6} c^{\prime 2}
    \right) \right] \right\} $
    &
    $\left. \left.
    + \frac{s_{\w} x_Z^{B^{\prime}}}{s^{\prime}c^{\prime}}
    \left( \frac{1}{5} - \frac{1}{2} c^{\prime 2} \right)
    \right] \right\}$ \\
\hline
$Z_L \bar d d$ &
    $-\frac{g}{2c_{\w}} \left\{ (-\frac{1}{2} + \frac{2}{3} s^2_{\w})
    - \frac{v^2}{f^2} \left[ -c_{\w} x_Z^{W^{\prime}} c/2s
    \right. \right.$ &
    $-\frac{g}{2c_{\w}} \left\{ \frac{1}{2}
    - \frac{v^2}{f^2} \left[ c_{\w} x_Z^{W^{\prime}} c/2s
    \right. \right.$ \\
$$ &
    $\left. \left.
    + \frac{s_{\w} x_Z^{B^{\prime}}}{s^{\prime}c^{\prime}}
    \left( 2y_u + \frac{11}{15} + \frac{1}{6} c^{\prime 2}
    \right) \right] \right\}$ &
    $\left. \left.
    + \frac{s_{\w} x_Z^{B^{\prime}}}{s^{\prime}c^{\prime}}
    \left( -\frac{1}{5} + \frac{1}{2} c^{\prime 2} \right)
    \right] \right\}$ \\
\hline
$A_H \bar u u$ &
    $\frac{g^{\prime}}{2s^{\prime}c^{\prime}}
    \left( 2y_u + \frac{17}{15} - \frac{5}{6} c^{\prime 2} \right)$ &
    $\frac{g^{\prime}}{2s^{\prime}c^{\prime}}
    \left( \frac{1}{5} - \frac{1}{2} c^{\prime 2} \right)$ \\
$A_H \bar d d$ &
    $\frac{g^{\prime}}{2s^{\prime}c^{\prime}}
    \left( 2y_u + \frac{11}{15} + \frac{1}{6} c^{\prime 2} \right)$ &
    $\frac{g^{\prime}}{2s^{\prime}c^{\prime}}
    \left( -\frac{1}{5} + \frac{1}{2} c^{\prime 2} \right)$ \\
\hline
$Z_H \bar u u$ & $gc/4s$ & $-gc/4s$ \\
$Z_H \bar d d$ & $-gc/4s$ & $gc/4s$ \\
\hline\end{tabular} \caption{Neutral gauge boson-fermion couplings
and $y_u = -2/5$ and $y_e = 3/5$ are required by the anomaly cancelation. The couplings are given
in the form $i \gamma^{\mu} (g_V + g_A
\gamma^5)$.} \label{gauge-fermion2}
\end{table}

\subsection{the $LH$ $ \phi W $ associated production at the LHC}
At the LHC, the parton level cross sections 
 are calculated at the leading order as
\begin{equation}
\hat{\sigma}(\hat s)
=\int_{\hat{t}_{min}}^{\hat{t}_{max}}\frac{1}{16\pi \hat{s}^2}
\overline{\Sigma}|M_{ren}|^{2}d\hat{t}\,,
\end{equation}
with
\begin{eqnarray}
\hat{t}_{max,min} =\frac{1}{2}\left\{ m_{p_1}^2 +m_{p_2}^2 -\hat{s}
\pm \sqrt{[\hat{s} -(m_{p_1}+m_{p_2})^2][\hat{s} -(m_{p_1}
-m_{p_2})^2]} \right\},
\end{eqnarray}
where $p_1$ and $p_2$ are the first and the second initial particles
in the parton level, respectively. For our case, they could be gluon
$g$ and quarks $u$, $d$, $c$, $s$, $b$ etc.

 The total hadronic cross section for $pp \to SS'+X$ can be
obtained by folding the subprocess cross section $\hat{\sigma}$ with
the parton luminosity
\begin{equation}
\sigma(s)=\int_{\tau_0}^1 \!d\tau\, \frac{dL}{d\tau}\, \hat\sigma
(\hat s=s\tau) ,
\end{equation}
where $\tau_0=(m_{p_1}+m_{p_2})^2/s$, and $s$ is the $p p$
center-of-mass energy squared. $dL/d\tau$ is the parton luminosity
given by
\begin{equation}
\frac{dL}{d\tau}=\int^1_{\tau} \frac{dx}{x}[f^p_{p_1}(x,Q)
f^{p}_{p_2}(\tau/x,Q)+(p_1\leftrightarrow p_2)],\label{dis-pp}
\end{equation}
where $f^p_{p_1}$ and $f^p_{p_2}$ are the parton $p_1$ and $p_2$
distribution functions in a proton, respectively.
 In our numerical
calculation, the CTEQ6L parton distribution function is
used~\cite{Ref:cteq6} and take factorization scale $Q$ and the
renormalization scale $\mu_F$ as $Q=\mu_F = m_{\phi} + m_W$. The loop
integrals are evaluated by the LoopTools
package~\cite{Ref:LoopTools}.

As for the SM parameters, throughout this paper, we take $m_t =173 $ GeV
\cite{topmass}, $m_W=80.38 $ GeV, $m_Z=91.19 $ GeV, and
$G_{F}=1.16637\times10^{-5}GeV^{-2}$ \cite{datagroup}, 
$\alpha_s(m_Z) = 0.118 $ and neglect bottom quark mass as well as
other light quark masses.

Now we discuss the main involved LH parameters,
\begin{itemize}
\item[{\rm (1)}]new scalar masses, which include
  the charged pseduo boson, neutral bosons and the SM-like higgs. The SM-like higgs
  has been discussed after the CERN experiment data release in e.g, Ref. \cite{1301.0090,1212.5930}, and the
  discussions show that the LH models may survive when $f\geq 800$ GeV.
     We here choose the loose constraints that $f\geq 500 $ GeV and
     the SM-like higgs mass as the current Experiment value: $125$
Gev \cite{atlas,cms}. The masses of other scalars $m_\phi$, despite of the small
electromagnetic difference, are the same and the constraints of them are
quite loose. 
  We here take $m_\phi$ as a free parameter varying from $200$ GeV to $600$ GeV,
  according to the references
such as Ref. \cite{LEP,higgs-lep,bg-h,th-han,1108.4416}.

 \item [{\rm (2)}] The mixing parameter s, c and s', c', which are
in the range of $0 \sim 1$. We will here take, however, $s$ free
parameter from $0-0.5$, and take $s'=0.5$ so as the  $c'>0.62$
according to Ref. \cite{litt-constr,0303236}.

 \item [{\rm (3)}] As for the scale $f$, one can have
a rough estimate of the natural scale  \cite{littlest}
\begin{equation}
    f \leq \frac{4 \pi m_H}{\sqrt{0.1 a_{\rm max}}}
    \simeq \frac{8 \ {\rm TeV}}{\sqrt{a_{\rm max}}}
    \left( \frac{m_H}{200 \ {\rm GeV}} \right),
\end{equation}
where $a_{\text{max}}$ denotes the largest coefficient 
 which could be of the order of $10$. So for a light
$m_H$, f may have a lower upper limit. 

We here also estimate $f$ in some particular situation, such as shown in
Ref. \cite{0303236,12032232}, in which the $U(1)$ sector can be modified by adding
an additional $U(1)$ and only gauging $U(1)_Y$, in the $SU(6)/SP(6)$ realization.
This may bring f to survive in the area of less than $1$ TeV.
In our calculation, we will weaken the constraints
a little and take $500< f< 2000$ GeV.
 \def\figsubcap#1{\par\noindent\centering\footnotesize(#1)}
 \begin{figure}[t]%
 \begin{center}
  \parbox{9cm}{\epsfig{figure=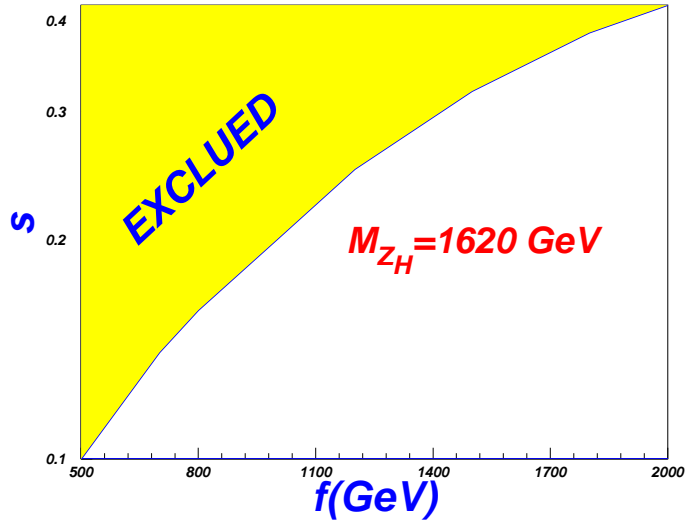,width=9cm}}
\caption{ In LH, the contour of the parameters $f$ and $s$ for the lower limit of the new
heavy gauge boson mass $M_{Z_H}=1.62$ TeV.
\label{con-f-s} }
 \end{center}
 \end{figure}

\begin{figure}[t]
\begin{center}
\epsfig{file=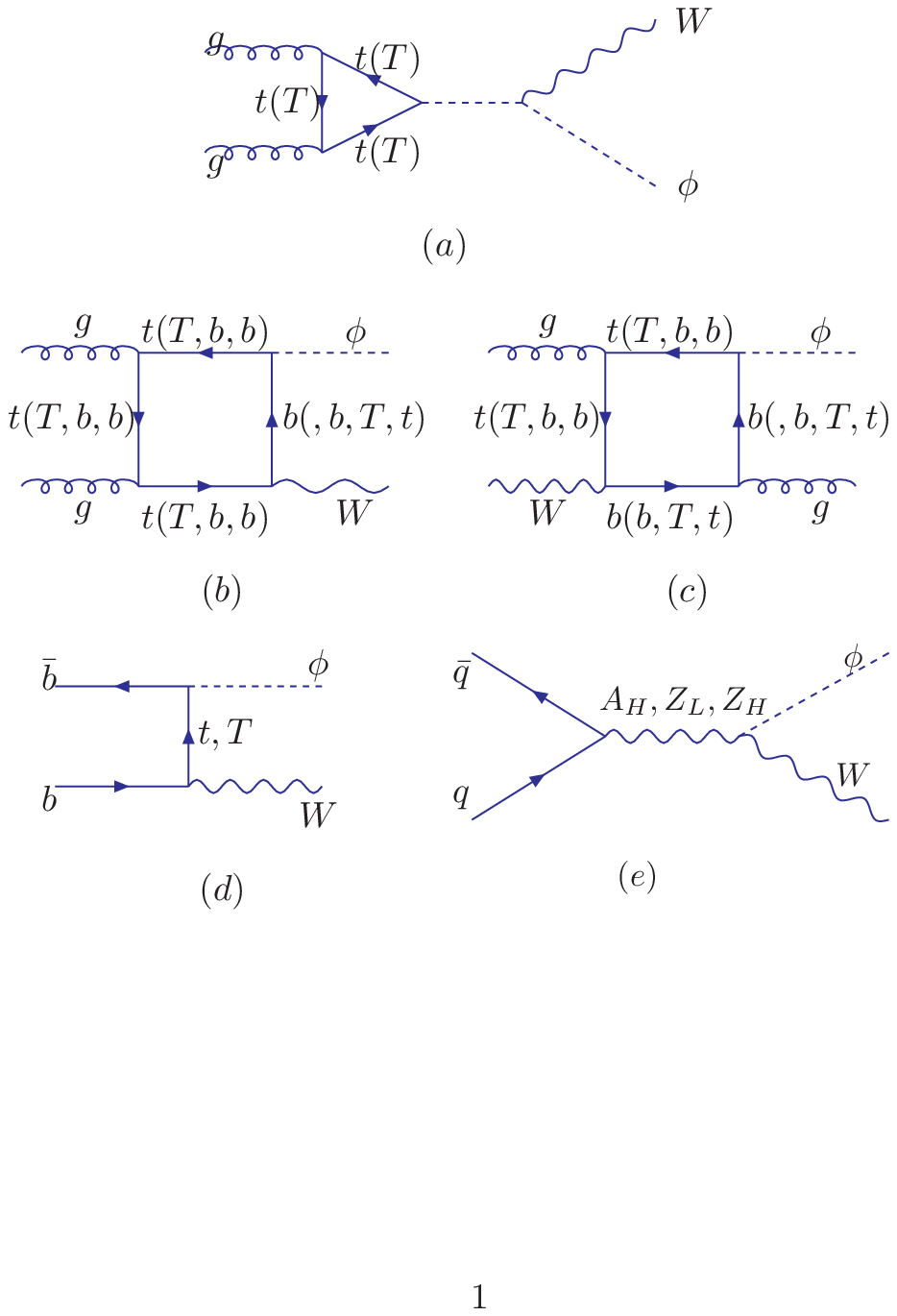,width=8cm}
\caption{Feynman diagrams for the charged scalar production
associated with the W boson at the LHC via gluon fusion and the
quark anti-quark annihilation parton level processes in the $LH$
model. Those obtained by exchanging the two external gluon lines are
not displayed here.} \label{ppsw-lh}
\end{center}
\end{figure}

 \item [{\rm (4)}]About the new gange boson masses, the charged and the neutral gauge bosons final masses
 related to our calculation are
written as\cite{littlest},
\begin{eqnarray}
    M_{W_L^{\pm}}^2 &=& m_w^2 \left[
    1 - \frac{v^2}{f^2} \left( \frac{1}{6}
    + \frac{1}{4} (c^2-s^2)^2
    \right) + 4 \frac{v^{\prime 2 }}{v^2}\right],
        \label{MWL}\\
    M_{Z_L}^2 &=& {m_z^2}
    \left[ 1 - \frac{v^2}{f^2} \left( \frac{1}{6}
    + \frac{1}{4} (c^2-s^2)^2
    + \frac{5}{4} (c^{\prime 2}-s^{\prime 2})^2 \right)+ 8
    \frac{v^{\prime 2 }}{v^2}  \right],
    \label{MZL}\\
    M_{A_H}^2 &=& m_z^2 s_\w^2 \left(
    \frac{ f^2 }{5 s^{\prime 2} c^{\prime 2}v^2}
    - 1 + \frac{x_H c_\w^2}{4s^2c^2  s_\w^2} \right)
    \label{MAH} \\
    M_{Z_H}^2 &=& m_w^2 \left( \frac{f^2}{s^2c^2 v^2}
    - 1 -  \frac{x_H s_\w^2}{s^{\prime 2}c^{\prime 2}c_\w^2}\right) ,
    \label{MZH}
\end{eqnarray}
where $m_z\equiv {gv}/(2c^{}_\w)$ and $m_w =m_zc_w$ are the SM neutral and charged boson mass, 
and the $x_H^{}$ can be found in Ref. \cite{littlest},
The masses of the 
$Z_H$, however, are still constrained by
 the LHC experiments. For example, the ATLAS\cite{1303.4287,1210.1718} and the CMS collaborations
  \cite{1302.4794,1307.1400} have detected the heavy vector boson as a di-jet resonance and give
the lower limits of the bosons, i.e, 
$M_{Z_H}>1.62$ TeV, which bound the limits of the
parameter involved. Since the last term of the $M_{Z_H}$ in Eq. \ref{MZH} is very small, the first term
decides the relation of the parameter $f$ and $s$, that is, the parameter $f$ and $s$ are restricted by
each other when the $Z_H$ mass range is set. we show in FIG.\ref{con-f-s} that the contour of the two parameters given the lower limit of the new heavy neutral mass $M_{Z_H}$.
Form FIG. \ref{con-f-s} we can see that the $f$ should be large for large $M_{Z_H}$, but if the $s$ becomes small, such as less than $0.15$ fb, $f$ may also be small in a narrow parameter space. This, however,
influence on our results is not too large since we take a quite a small $s$, for example, $s=0.1$, in most of calculation unless specifically stated.


\item [{\rm (5)}] If we define $x=4fv'/v^2$, where $v'$ is the vacuum expectation value( VEV)
of the scalar of the triplet $\phi$, the masses of the neutral boson in the above equations
can be rewritten by the parameter $x$ and
by this definition, the neutral scalar mass can be given as,
\begin{equation}
M^{2}_{\phi^{0}}=\frac{2m^{2}_{H^{0}}f^{2}}{v^{2}[1-(4v'f/v^{2})^{2}]}=\frac{2m^{2}_{H^{0}}f^{2}}{v^{2}(1-x^{2})}
\end{equation}

The above equation about the mass of $\phi^0$ requires a constraint
of $0\leq x<1$ ( i.e.,$4v'f/v^2<1$ ), which shows the relation
between the scale $f$ and the VEV of the higgs field doublets and
the triplet ($v,~v'$),
but this constraint is quite loose, in which the $v'$ can be as large as $20-30$ GeV for a small $f$ value.
The parameter $\rho\equiv m_zc_w/m_w$ (also the $T$ parameter) dependence on the $v'$, however, has been
studied in Ref. \cite{litt-constr}  and find that in a quite a large space, the $v'$
may lie in the range of several GeV, which constraints the parameter  $x$ can not be too
large. This is also the constraint coming from $T$ parameter since $T$ can be written as $\alpha T=\rho-1=\Delta \rho$.
With the constraint of $v'$ in the order of several GeV, we can here take $x$ as a free parameter
in the range $0<x<0.2$, which also indicate clearly that more larger $x$ is
not allowed by current experiments.

\item [{\rm (6)}] In the LH model, the relation among the Fermi
coupling constant $G_{F}$, the gauge boson W mass $M_{W}$ and the
fine structure constant $\alpha$
 can be written as\cite{gf,littlest}:
\begin{eqnarray}
\frac{G_{F}}{\sqrt{2}}=\frac{\pi\alpha}{2M^{2}_{W}s^{2}_{W}}[1-c^{2}(c^{2}-s^{2})\frac{v^{2}}{f^{2}}
+2c^{4}\frac{v^{2}}{f^{2}}-\frac{5}{4}(c'^{2}-s'^{2})\frac{v^{2}}{f^{2}}]
 \end{eqnarray}
 So we have
\begin{eqnarray}
\frac{e^{2}}{s^{2}_{W}}=\frac{4\sqrt{2}G_{F}M^{2}_{W}}{[1-c^{2}(c^{2}-s^{2})\frac{v^{2}}
{f^{2}}+2c^{4}\frac{v^{2}}{f^{2}}-\frac{5}{4}(c'^{2}-s'^{2})\frac{v^{2}}{f^{2}}]}
 \end{eqnarray}
\item [{\rm (7)}] Finally, the recent data of the $125$ GeV higgs also
put some constraints on the parameter space \cite{1301.0090}, but the constraints are quite loose,
for example, FIG.1 and FIG.2 in Ref. \cite{1301.0090}
give the dependence of the ratio $R$ ($R=Br(h\to \gamma\gamma(ZZ))_{LH}/Br(h\to \gamma\gamma(ZZ))_{SM}$)
 on the f and find that they put quite loose constraints to the parameters, so we will not discuss
 further.
\end{itemize}

\def\figsubcap#1{\par\noindent\centering\footnotesize(#1)}
 \begin{figure}[htb]%
 \begin{center}
  \parbox{7cm}{\epsfig{figure=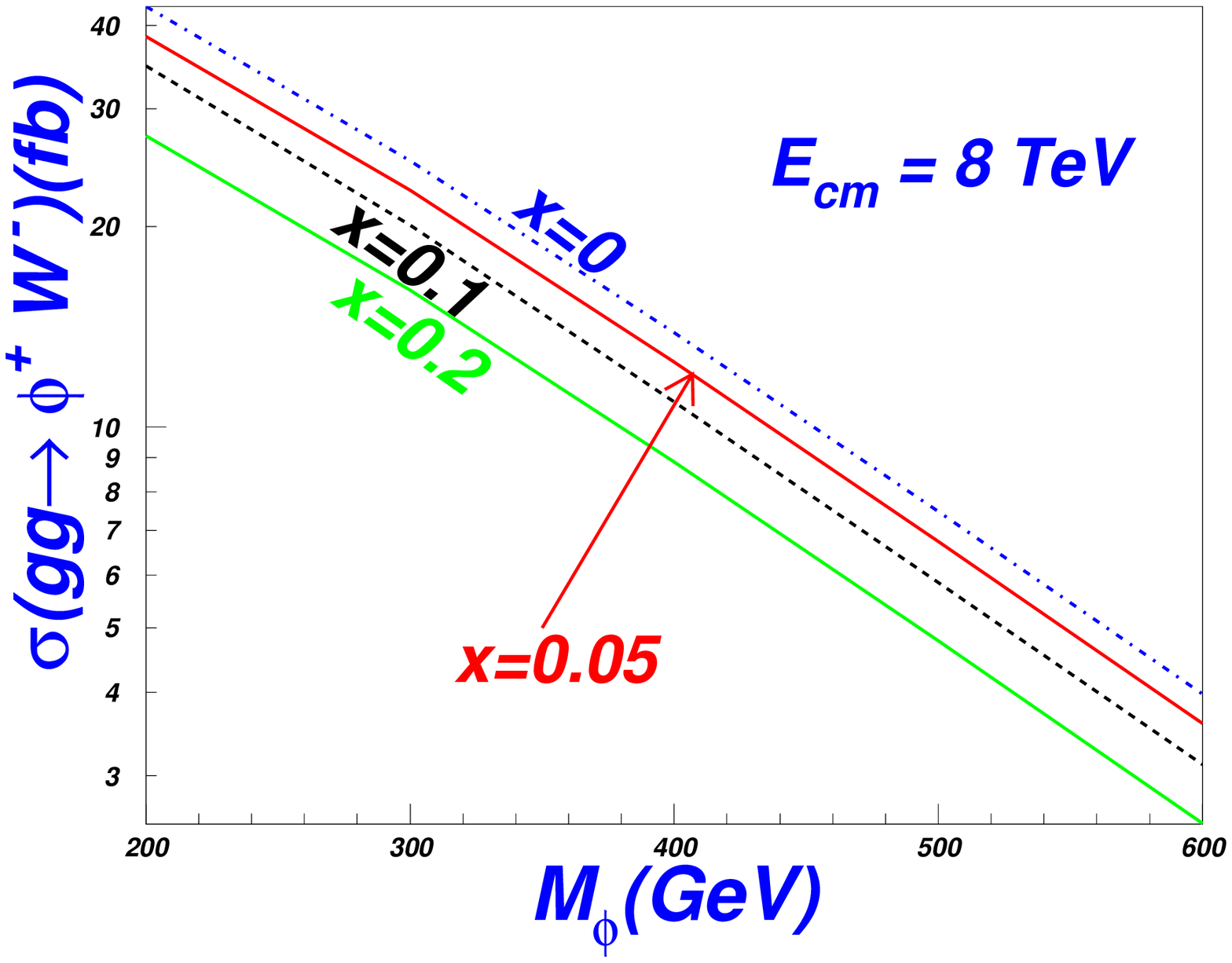,width=7cm}    \figsubcap{a}}
  \parbox{7cm}{\epsfig{figure=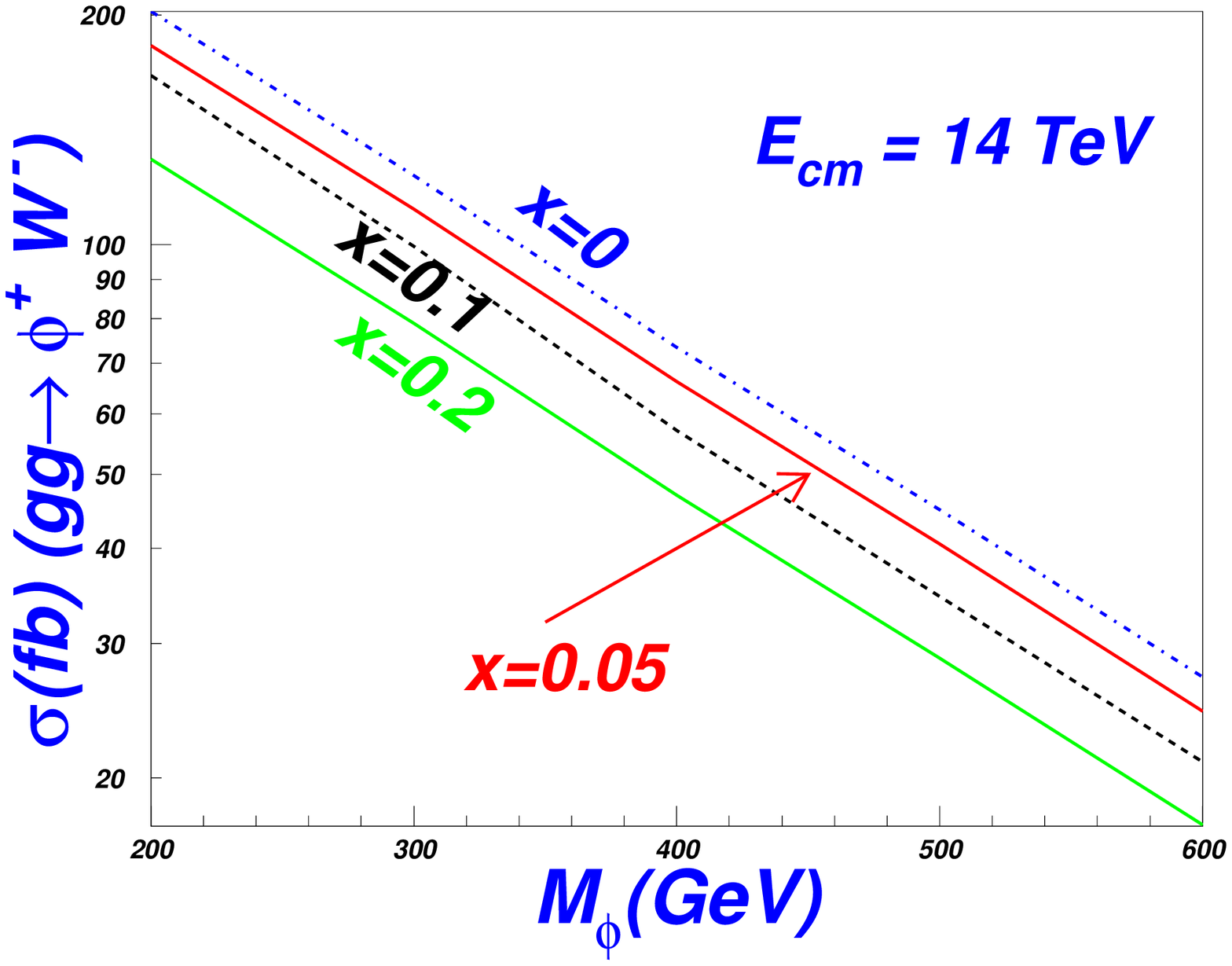,width=7cm}    \figsubcap{b}}
  \parbox{7cm}{\epsfig{figure=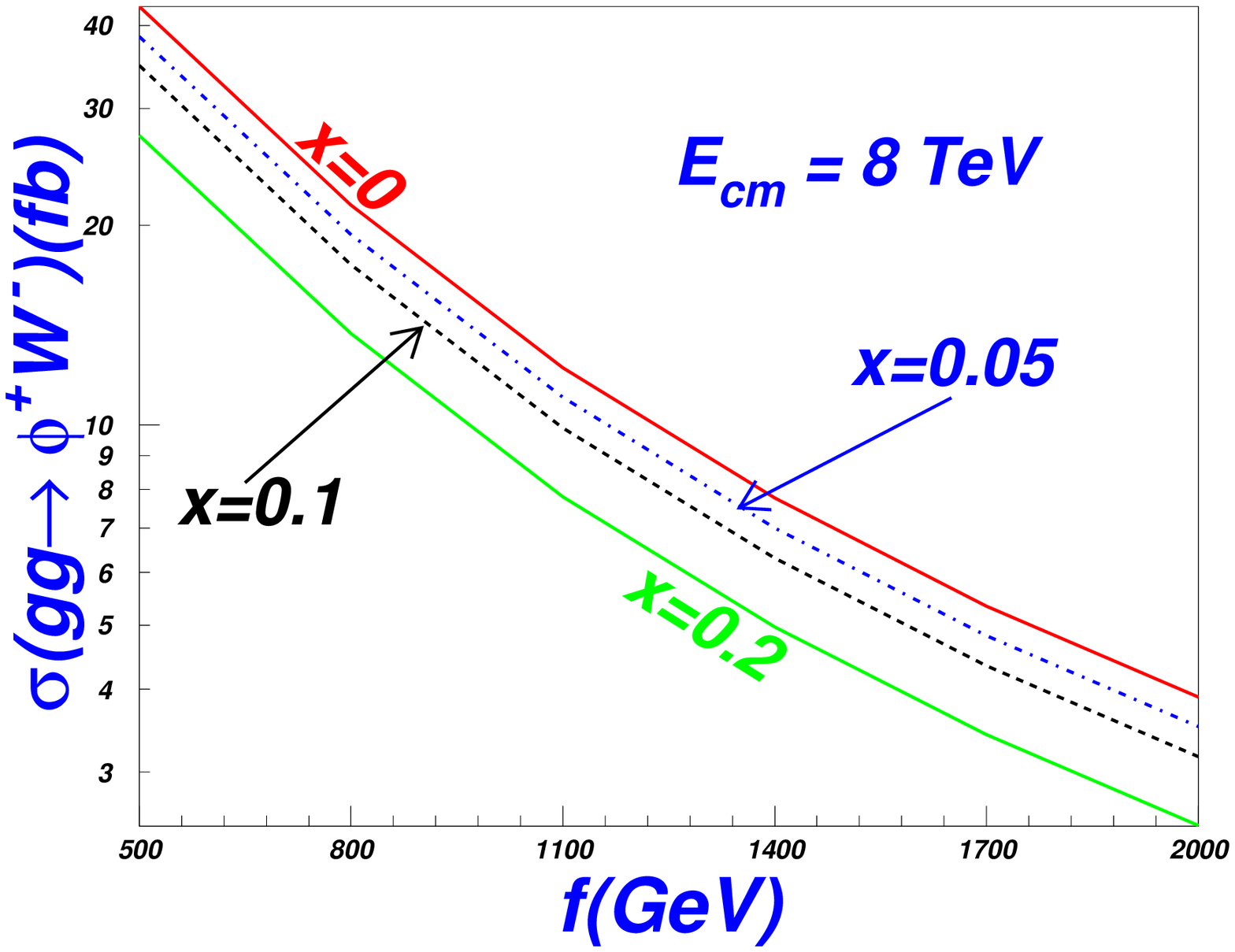,width=7cm}   \figsubcap{c}}
  \parbox{7cm}{\epsfig{figure=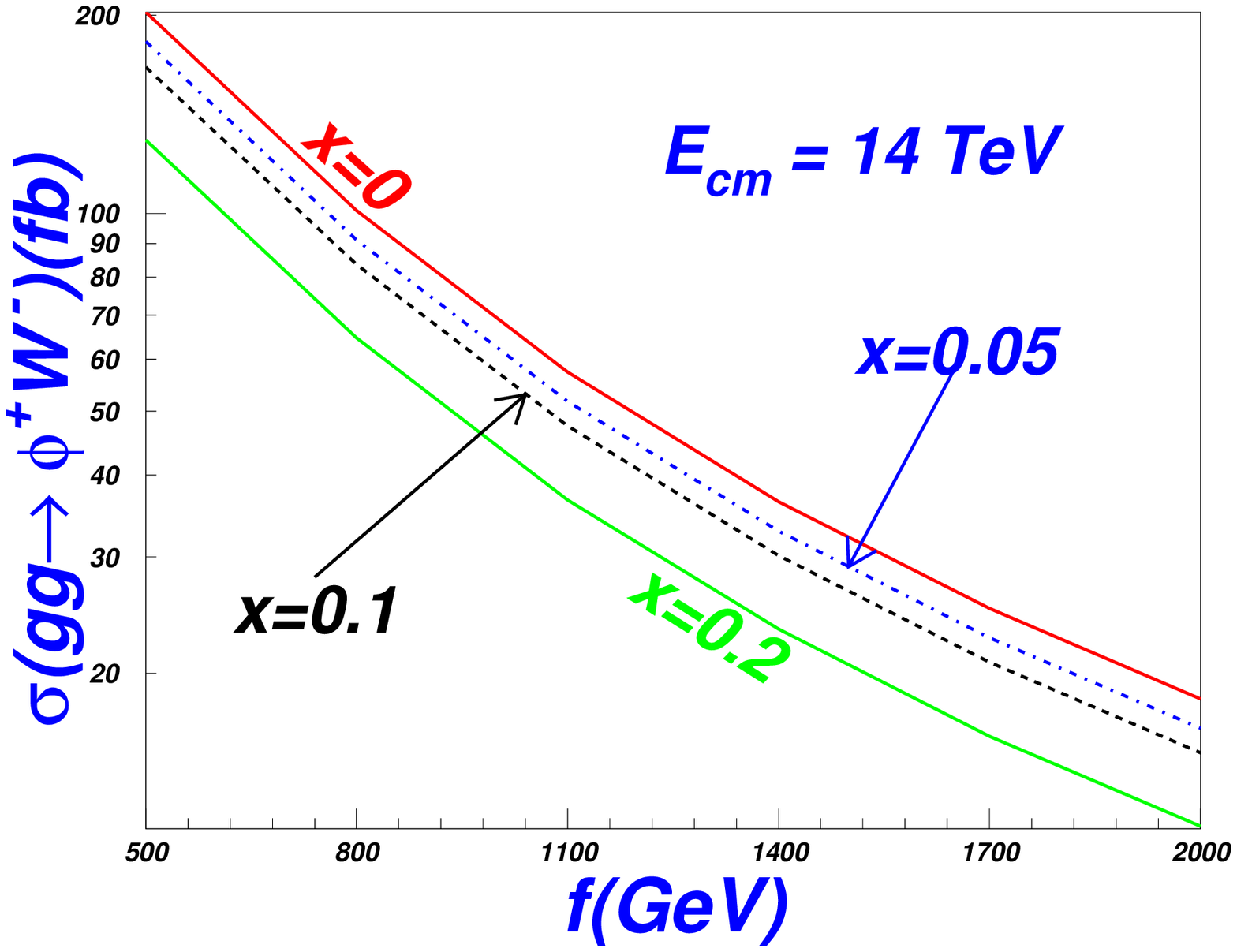,width=7cm}    \figsubcap{d} }
 \caption{ In LH, the cross section $\sigma$ of the processes $gg \to \phi^+W^-$
as a function of the scalar mass $m_{\phi}$ with the center-of-mass energy $E_{cm}=8$ TeV
(a) and $14$ TeV (b)  respectively, for $f=500$ GeV  and
$s=~0.1$, with different x $x= 0, ~0.05,~0.1,~0.2$; the cross
section $\sigma$ of the processes $gg \to \phi^+W^-$ as a function
of $f$ with $E_{cm}=8$ TeV (c) and $14$ TeV (d)
respectively, and $s=~0.1$, with different $x$ ($x= 0, ~0.05,~0.1,~0.2$). \label{ggsw-1} }
 \end{center}
 \end{figure}
 \begin{figure}[htb]%
  \begin{center}
 \parbox{5.5cm}{\epsfig{figure=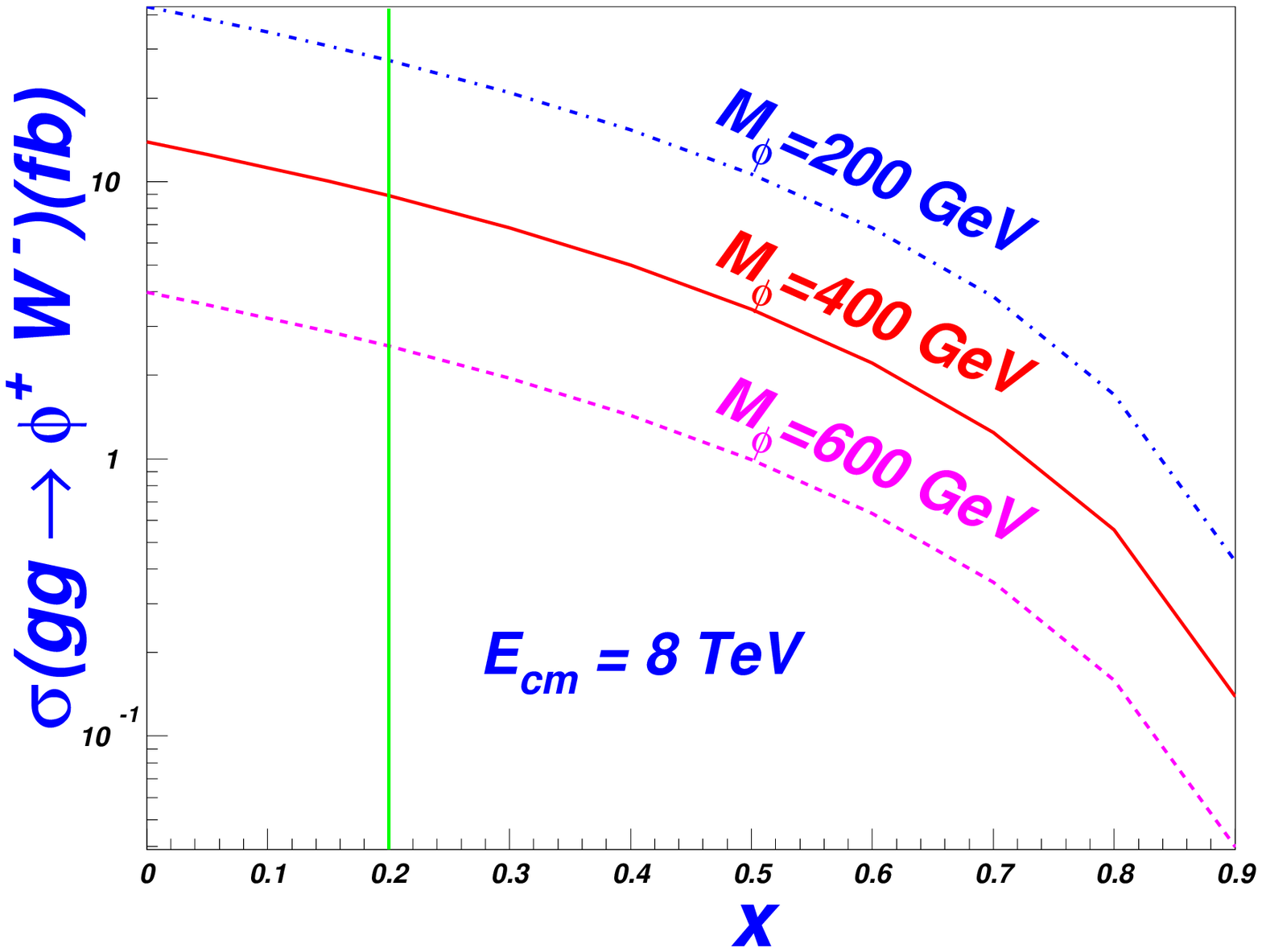,width=5.5cm}   \figsubcap{a}}
 \parbox{5.5cm}{\epsfig{figure=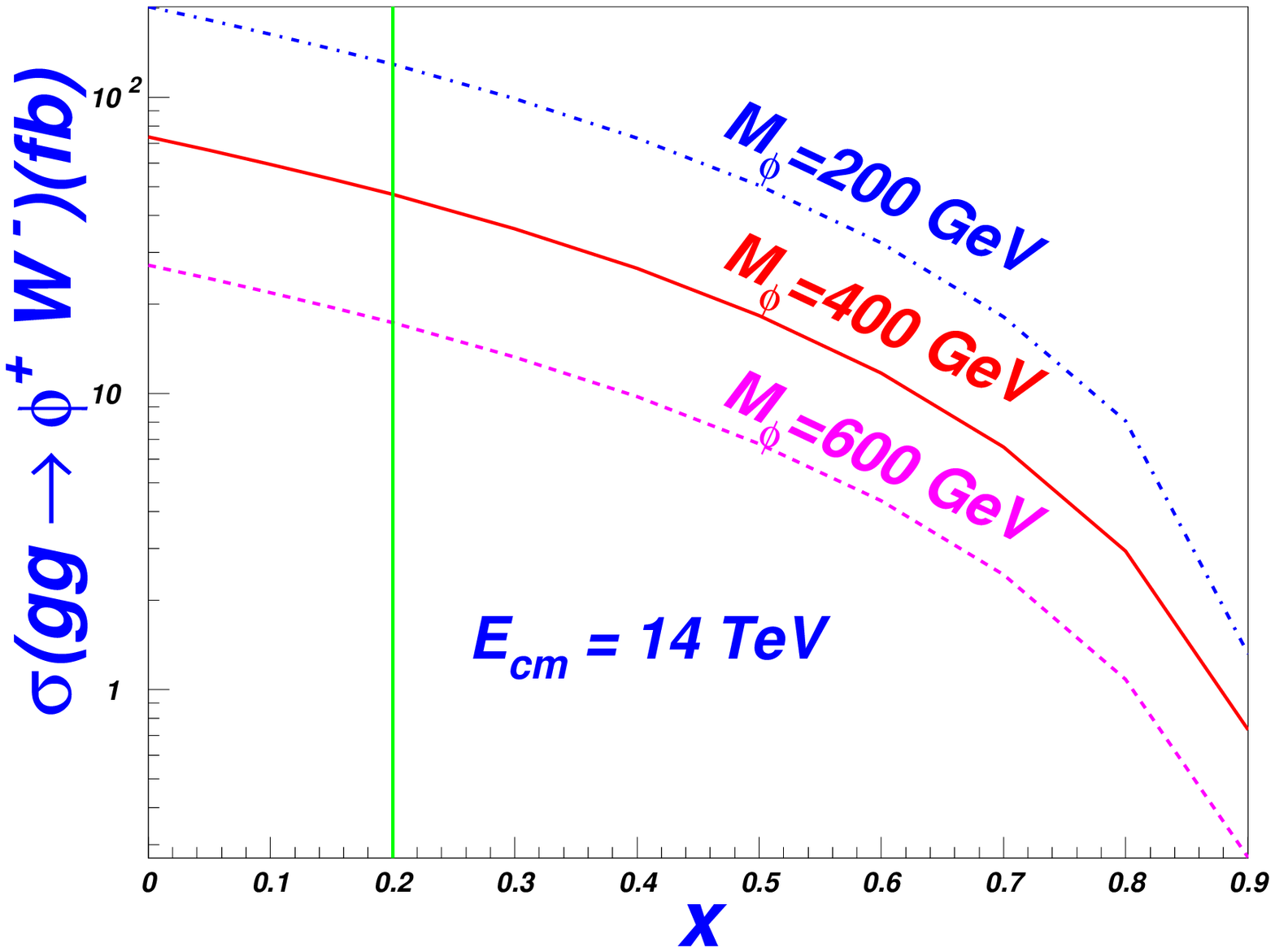,width=5.5cm}    \figsubcap{b} }
 \parbox{5.5cm}{\epsfig{figure=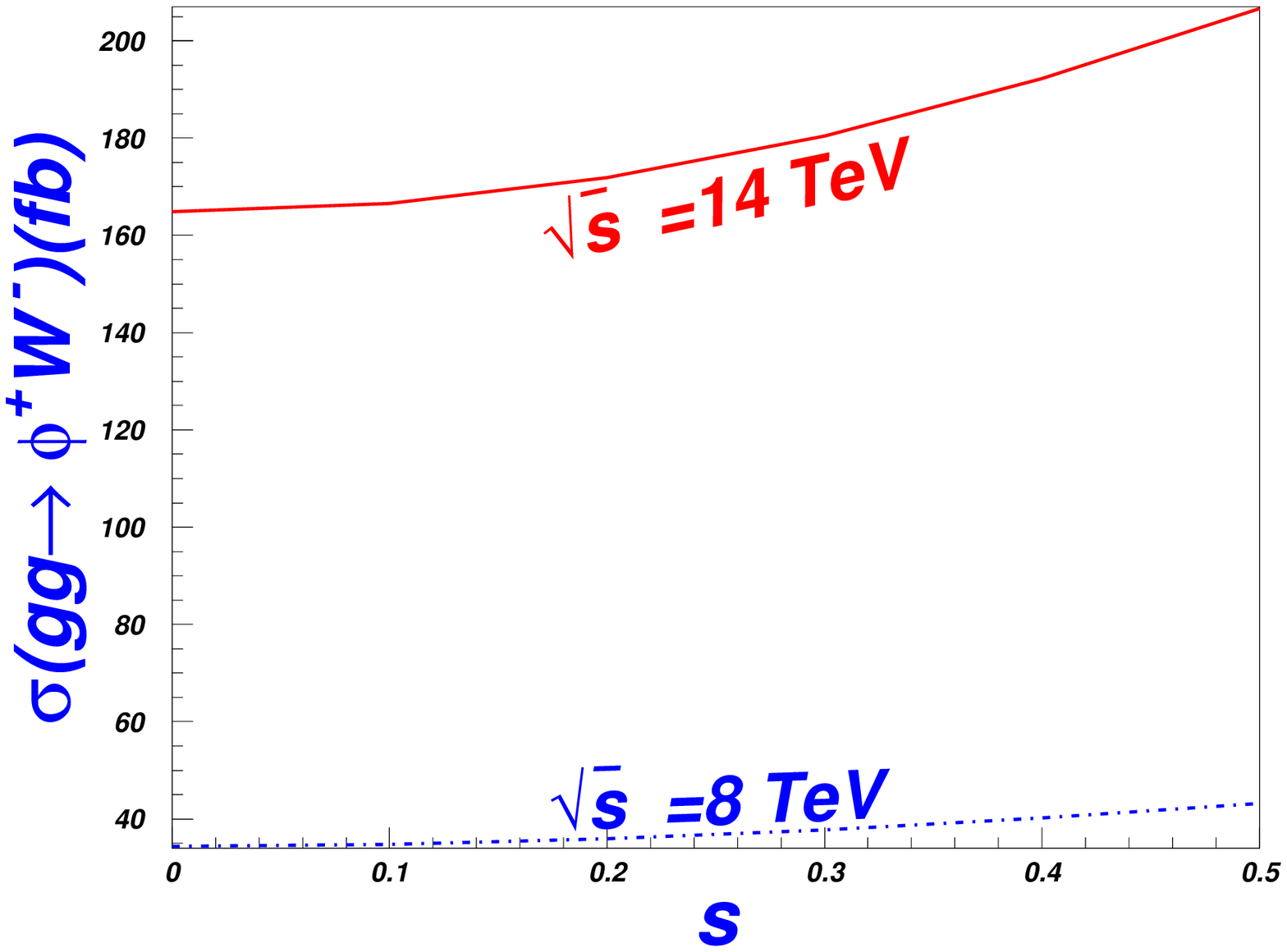,width=5.5cm}    \figsubcap{c}}
  \caption{In LH, when the scalar mass $m_{\phi}=200,~400,~600$ GeV, the cross section $\sigma$
of the processes $gg \to \phi^+W^-$ as a function of $x$ with
$E_{cm}=8$ TeV (a) and $14$ TeV (b) respectively, and
$s=~0.1$,  with $f=500$ GeV; The cross section $\sigma$ of the
processes $gg \to \phi^+W^-$ as a function of $s$ for $x=0.1$,
$f=500$ GeV and $E_{cm}=8$ TeV
and $14$ TeV (c).
\label{ggsw-2} }
 \end{center}
 \end{figure}
\subsection{numerical results in littlest higgs model}
 Due to the interactions in Tables \ref{gauge-scalar-fermion} and
\ref{gauge-fermion2}, the single charged boson production associated
with the W boson processes can proceed  through various parton
processes at the LHC, as shown in FIG. (\ref{ppsw-lh}), in which those
obtained by exchanging the two external gluon lines are not
displayed here. To know the relative values of them, we here,
firstly, discuss the contributions from every single parton channel
though , actually, we can not distinguish the initial states, i.e,
we will firstly discuss the $gg$ fusion and the $q\bar q$
annihilation processes, respectively, and then sum them all together
to see the total contributions.

\subsubsection{gg fusion in the LH models}

 \begin{figure}[htb]%
 \begin{center}
   \parbox{6cm}{\epsfig{figure=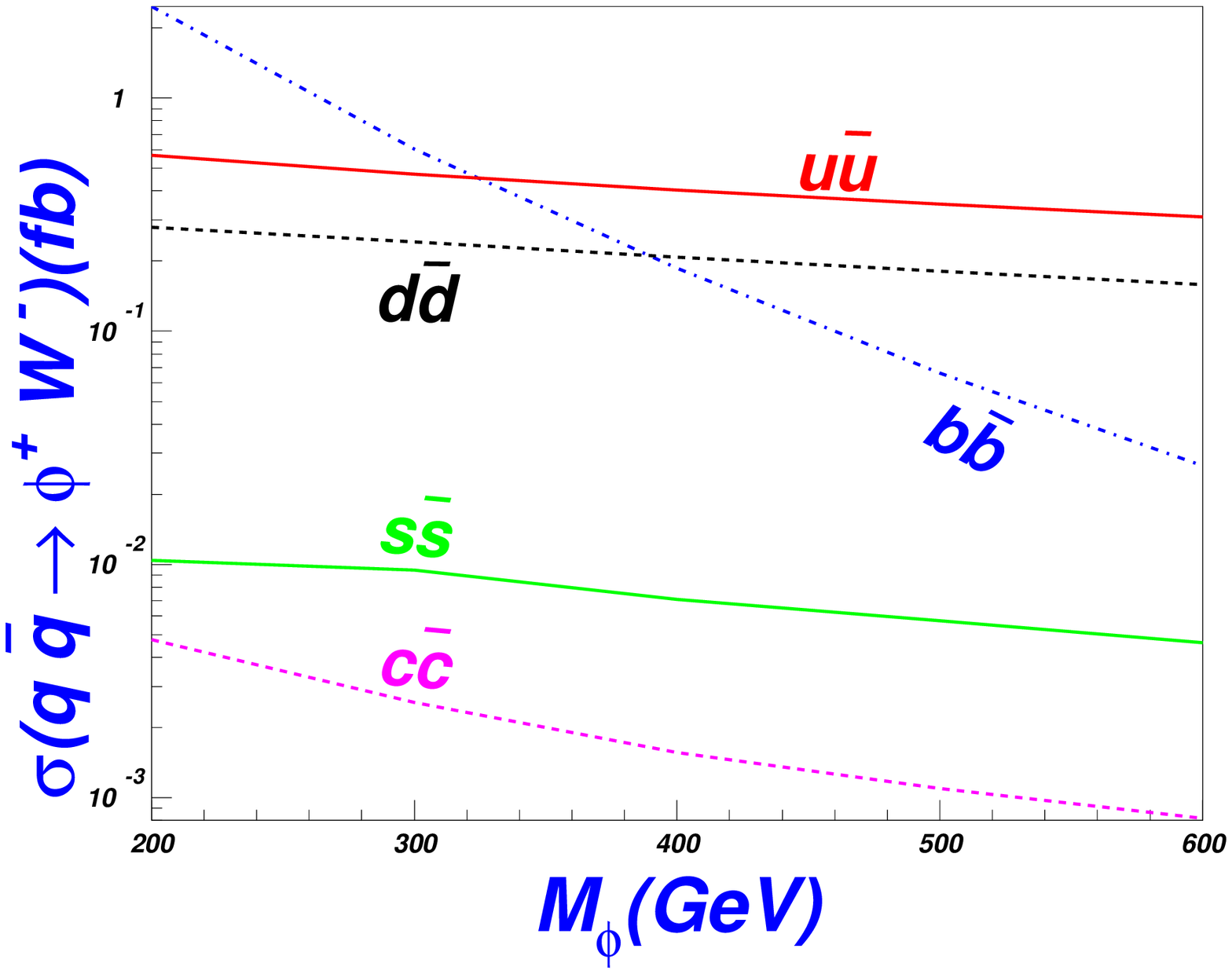,width=7cm}    \figsubcap{a}}
   \parbox{6cm}{\epsfig{figure=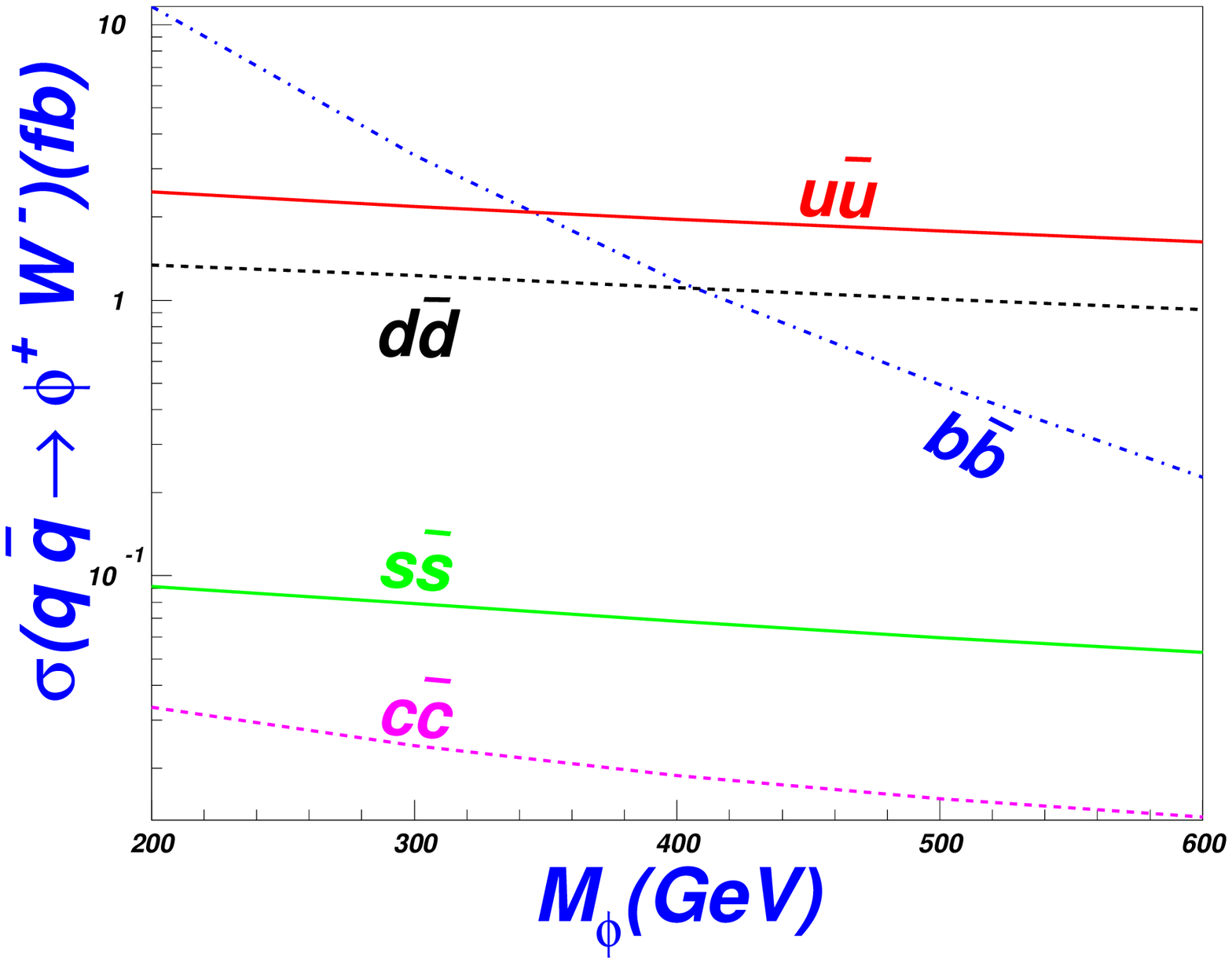,width=7cm}    \figsubcap{b}}
   \parbox{6cm}{\epsfig{figure=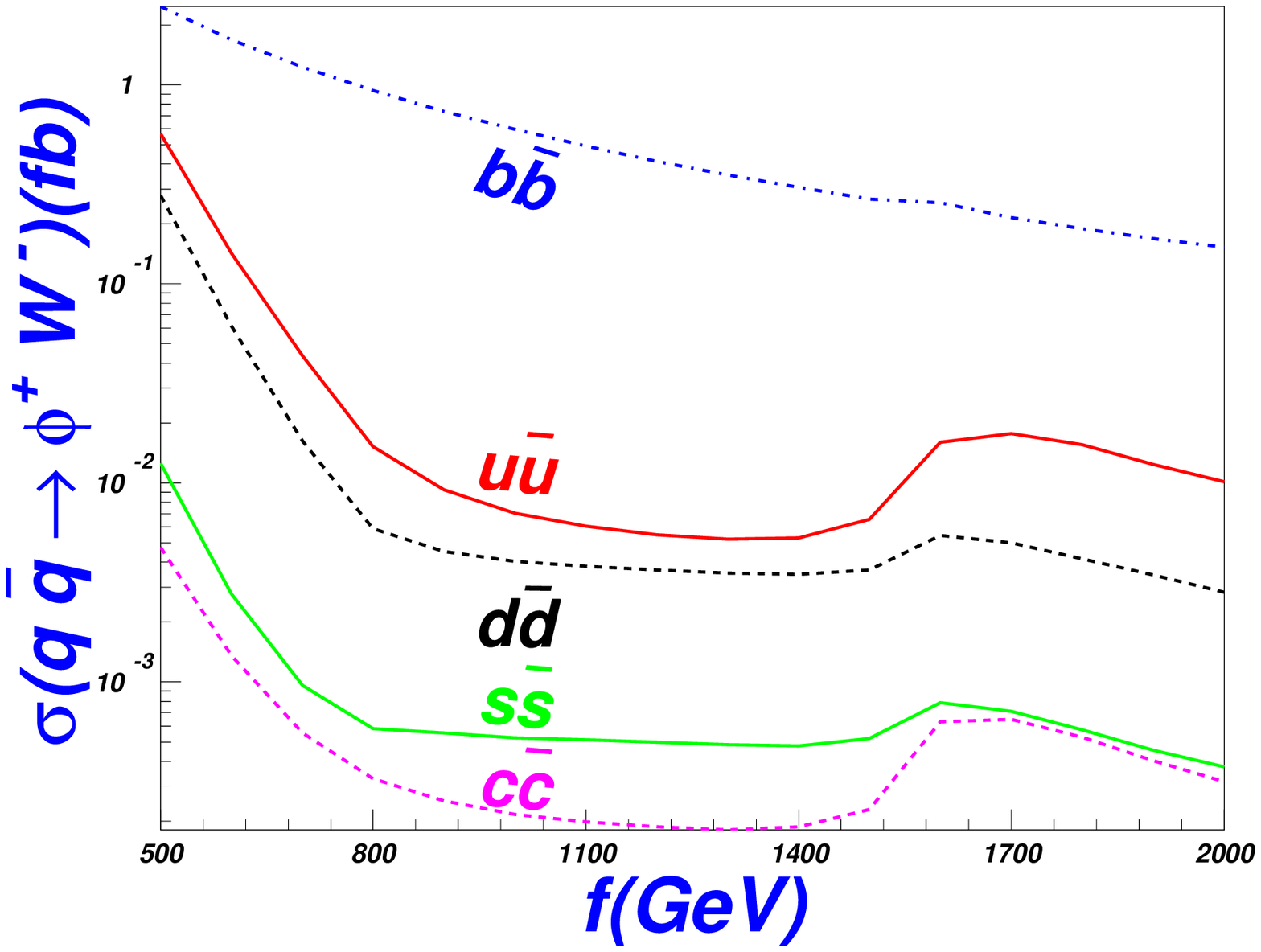,width=7cm}  \figsubcap{c}}
   \parbox{6cm}{\epsfig{figure=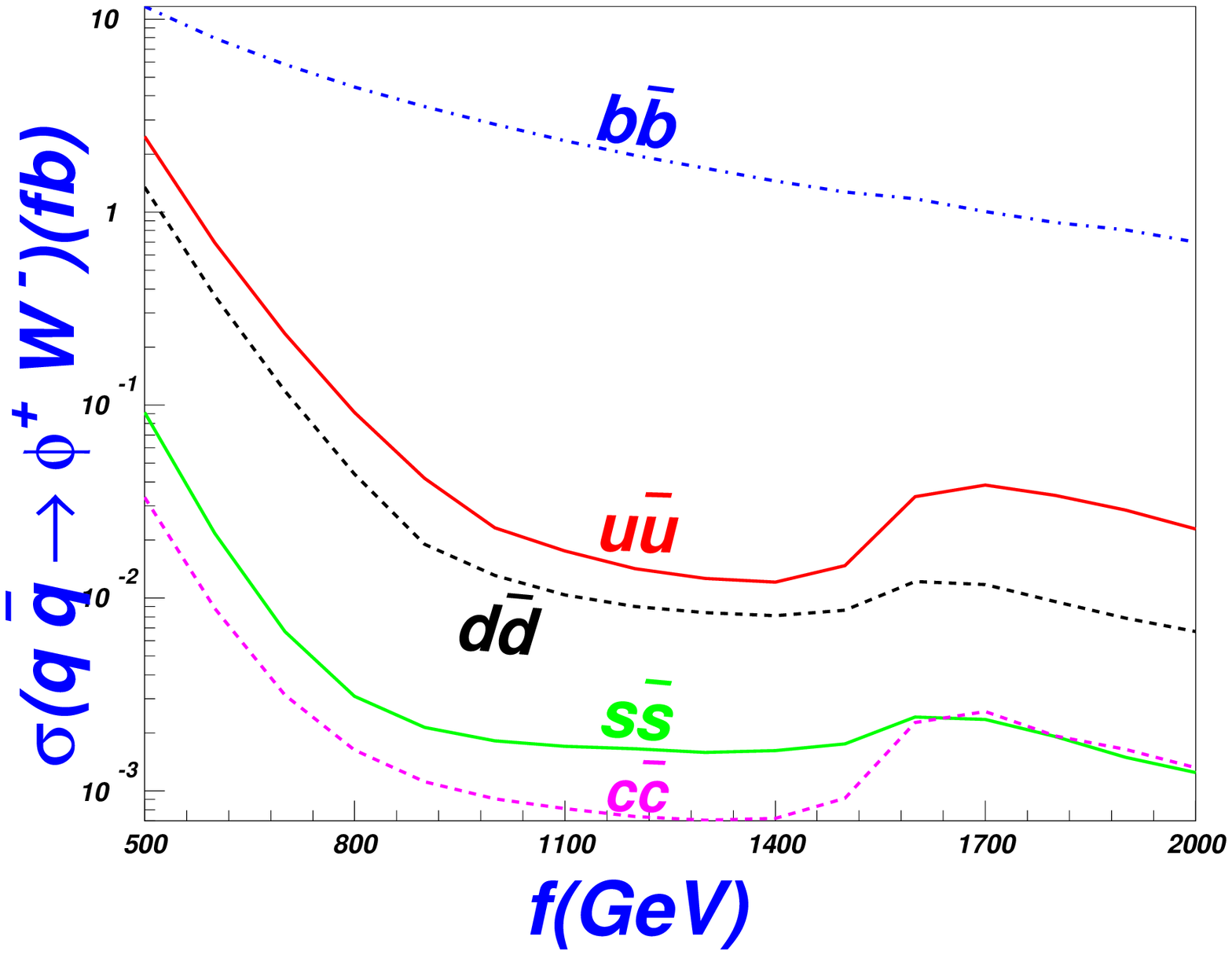,width=7cm}  \figsubcap{d}}
 \caption{ In LH, the cross section $\sigma$ of the processes $q\bar q \to \phi^+W^-$
  as a function of the scalar mass $m_{\phi}$ with $E_{cm}=8$ TeV (a) and $14$ TeV (b)
for $f=500$ GeV, $x=0.1$ and $s=0.1$; The $q\bar q \to \phi^+W^-$ cross
section $\sigma$ as a function of $f$ with $E_{cm}=8$ TeV (c) and $14$ TeV (d)
respectively, and $s=~0.1$, $x=0.1$. Here $q=u,d,c,s,b$ quarks. \label{qqsw-1} }
 \end{center}
 \end{figure}
\def\figsubcap#1{\par\noindent\centering\footnotesize(#1)}
 \begin{figure}[b]%
 \begin{center}
  \parbox{6cm}{\epsfig{figure=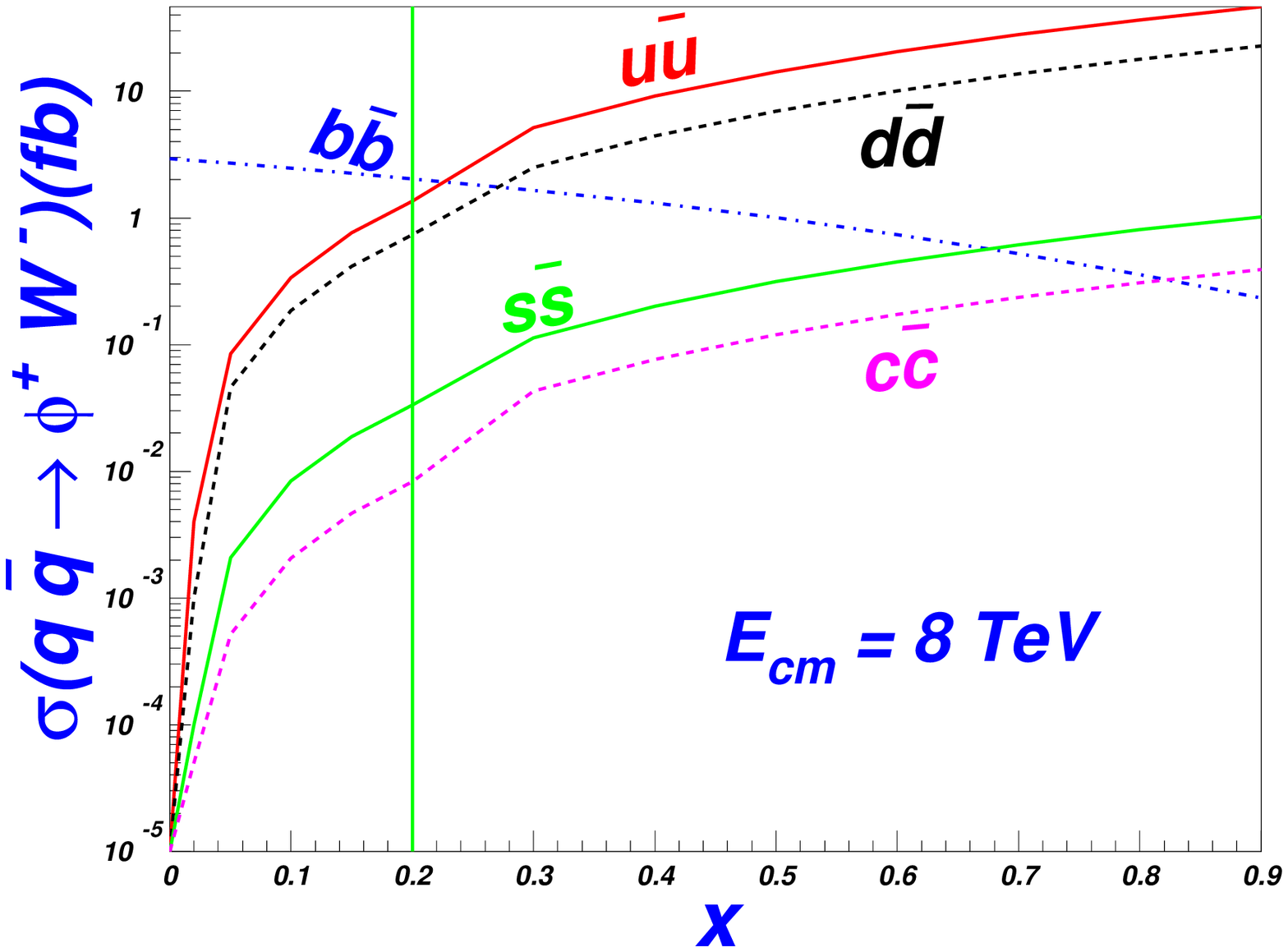,width=6cm}  \figsubcap{a}}
  \parbox{6cm}{\epsfig{figure=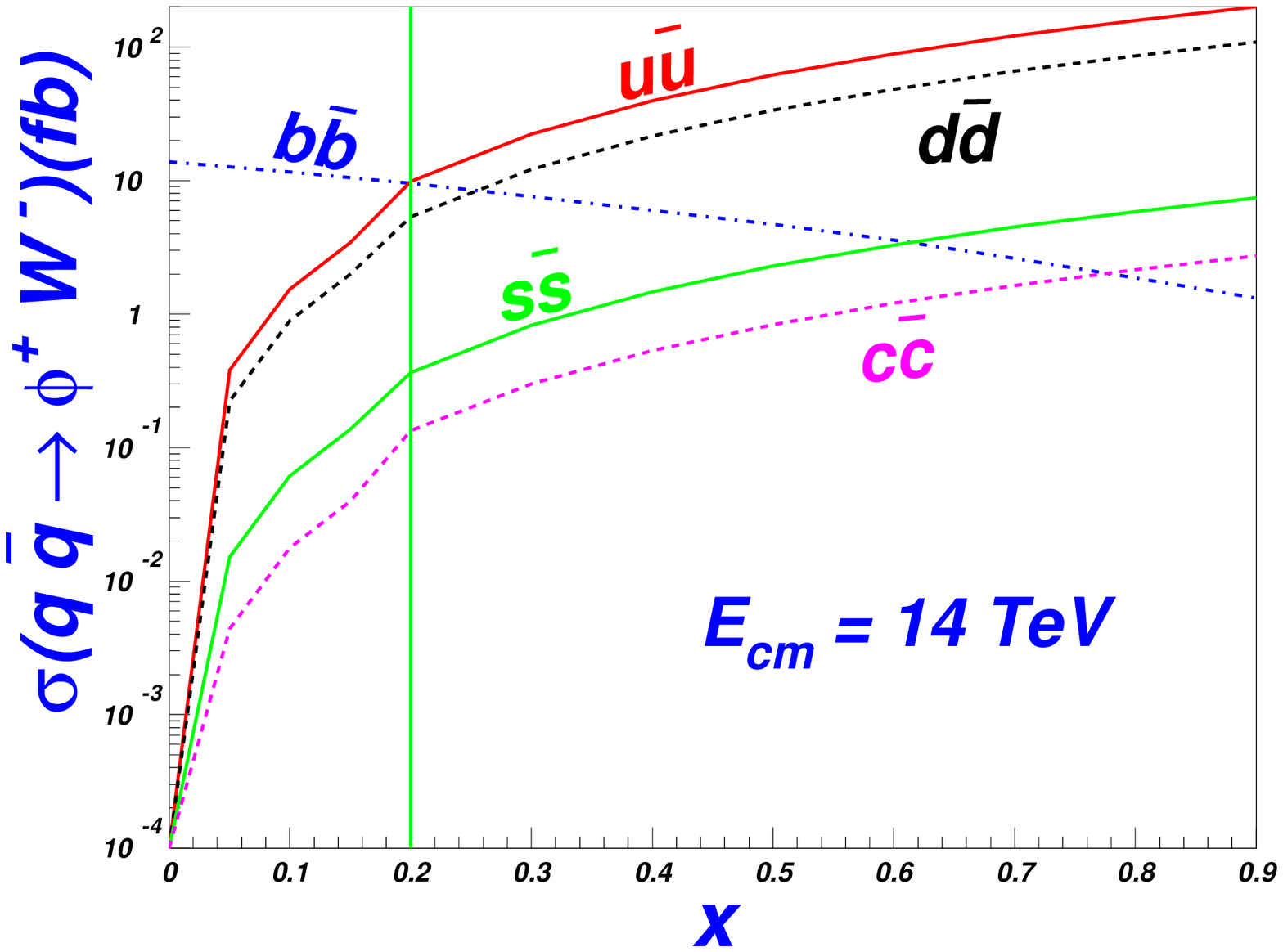,width=6cm}  \figsubcap{b}}
 \caption{ In LH, the cross section $\sigma$ of the processes $q\bar q \to \phi^+W^-$,
with $m_{\phi}=200$ GeV, 
$f=500$ GeV£¬  as a function of $x$ 
for  $E_{cm}=8 $ TeV (a) and
$14$ TeV (b) ($s=0.1$). 
\label{qqsw-2} }
 \end{center}
 \end{figure}
Note that the processes consist of the box diagrams and the
$W$ scalar scalar coupling, just shown as FIG. (\ref{ppsw-lh}) (a) and
(b)(c), The s-channel contribution of the cross sections, however,
is tiny, which is easy to understand with the quite large
center-of-mass suppression.

The production cross sections of the $\phi^+ W^-$ of the $gg$ fusion
are plotted in FIG. (\ref{ggsw-1}) for $E_{cm}=8,~14$ TeV,
respectively, with $x=0,~0.05,~0.1,~0.2$ and $f=500$ GeV, as
functions of the scalar mass $m_\phi$, assuming the charged and
neutral scalar mass degenerate, $m_{\phi^\pm}
=m_{\phi^0}=m_{\phi^p}$. From FIG. (\ref{ggsw-1}), we can see the
cross section of this process is quite large, about $100$ fb in most
of the parameter space and, as was expected, the production rate
decreases with the increasing scalar mass since the phase space are
suppressed by the final masses.

To compare the other parameter dependence, in FIG. (\ref{ggsw-1}) (c) (d) we
give the cross sections depend on the parameter $f$ and $s$, for
$E_{cm}=8,~14$ TeV, $f=500$ GeV, and $m_\phi = 200$ GeV, which can
clearly show the production rate varying as the different parameter.
We can see the increasing production rate with the increasing $s$,
but the cross section is decreasing when $f$ grows up.

FIG. (\ref{ggsw-2}) shows the parameter $x$ dependence of the cross sections, forgetting
the experimental constraints temporally, taking its value from $0$ to $0.9$ instead.
From both FIG. (\ref{ggsw-1}) and FIG. (\ref{ggsw-2}) we can also see the $x$
dependence of the process $gg\to \phi^+W^-$ is strong since the
$x~ ( =4fv'/v^2)$ is closely connected to the triplet VEV $v'$, and
the $v'$ decide the mixing parameter $s_+$, the parameter involved
in the $\phi^+ \bar t(T) b$. The production cross sections of the
processes $gg\to W^+\phi^-+X$ decrease with the increasing parameter
$x$. For example, when the center-of-mass is $8$ TeV, for
$x=0$, $m_\phi =200 $ GeV, and $s=0.1$,  the production rate is about $42$
fb; When  $x=0.2$, however, the production rate declines to only
$27$ fb.  The larger the $x$ is, the smaller the cross section is. The situation
is the same even $x$ increasing to $1$, though the
experiments constrain this parameter far below $1$.
When the center-of-mass is $14$ TeV, the same situation occurs, just
the rate will be about an order larger than those of the smaller
center-of-mass, i.e, $8$ TeV.

As we have discussed, too large $x$ is excluded by the current data, so a
vertical line is added in FIG.\ref{ggsw-2},
and in the left of the line is the allowed areas, while the right of it is
 disallowed. The same situations occur in
 FIG. \ref{qqsw-2} and FIG.\ref{ppsw-2}.

 The mixing $s$ affects the process $gg\to \phi^+W^-$ largely, too,
however, the trend is different. We can see from FIG. (\ref{ggsw-2})
(c) that the possibility of the $\phi^+W^-$ associated production
increases with increasing $s$. In FIGs. (\ref{ggsw-1}),
(\ref{ggsw-2})(a)(b), we take $s=0.1$, which is quite small compared
to its maximum value, $0.5$, according to the discussion of Ref.
\cite{litt-constr,0303236}. So our discussion are not the maximum, the results are
general.


Similarly, we can see from FIG. (\ref{ggsw-1})(c)(d) that the process
$gg\to \phi^+W^-$ is dependent strongly on the parameter $f$, which
is understandable since, the most couplings in the LH models, such
as $\phi^+ \bar t(T) b$ and $\phi^+ W^- S $ ($S=\phi^0,~\phi^p,~H$),
etc, are tightly connected with the parameter $f$. The cross
sections may be large if the scale $f$ is not too high and decreases
rapidly as the increasing $f$. The rates of the $\phi^+W^-$
production for $\sqrt s = 14 $ TeV, for example, can arrive at about
$273$ fb,  $35$ fb and $24$ fb, for $f=500$,  $1000$ and $2000$ GeV,
respectively, with $s=0.1$.

\subsubsection{$\phi W$ production via quark anti-quark annihilation }

 In LH, the $\phi W$ production via quark anti-quark annihilation are
realized by the parton level $u\bar u, ~d\bar d, ~c\bar c, ~s\bar s,
~b\bar b $ $ \to  \phi W$, which can be distinguished as t-channel
and s-channel processes, but the t-channel are only realized by the
$b\bar b$ initial state, via $\phi^+ \bar t(T) b$ couplings.

And for the s-channel scalar and $W$ boson associated production
induced by the $q \bar q$ ($q=u,d,s,c,b$) collision, what makes the
difference among them is only, if we neglect the masses of the
quarks $u,d,c,s, b$, the parton distribution function in the proton,
so it is naturally to see in FIG. (\ref{qqsw-1}) that $\sigma(u\bar u)
>\sigma(d\bar d) >\sigma(s\bar s)>\sigma(c\bar c)
>\sigma(b\bar b) $.

In FIGs.(\ref{qqsw-1}) (a)(b), we can see the cross sections decrease
with increasing charged scalar mass $m_\phi$, the level of decline,
however, is different. For $b\bar b$ realization, we can see it
declines rapidly, while the $u\bar u, ~\ddbar,~\ssbar,~\ccbar$
annihilations are not so quickly. That can be understood that
with the small distribution in the proton,  the $\bbbar$ collisions mainly
contribute via the t-channel, while the others are from s-channels
mediated by the bosons $A_H,~Z_L,~Z_H$ which appear in the  propagator,
with large masses about $1$ TeV, which weakens the effect of the increasing
scalar mass. When $m_\phi$ is not too large compared to the heavy boson mass,
the cross sections from s-channel $q\bar q$ annihilations
are almost unchanged. Actually, if we assume the $\phi^+$ mass are
in the order of the heavy bosons, i.e, more than $1$ TeV, the
situation is different immediately. With the increasing $m_\phi$,
the production rates decrease largely, which is verified by our calculation though
not shown here.

 The s-channel processes in FIG.~(\ref{ppsw-lh}) (e), though the parton
distribution functions are larger for the $u\bar u$ and $d\bar
d$ initial states, may be relatively small in view of the
center-of-mass suppression effects. At the same time, the t-channel
coupling strengths may be large for little $x$.
In FIG.~(\ref{ppsw-lh})(d), for instance, the
strength of $\phi^+ \bar t(\bar T) b \sim m_t/v \sim 1$ contributes large.
so no wonder the
cross sections of the parton level processes like $u\bar u(d\bar d,
s \bar s) \to \phi W $ are  smaller
than those of the others even with larger parton distribution
functions, especially with the increasing $f$. These can be seen clearly in
FIGs. (\ref{qqsw-1})(c)(d). 

Note that in FIG.(\ref{qqsw-1}) the processes depend largely on the
parameter $f$, and if the parameter $f$ decreases, the production
rate of this process will go up rapidly. From the couplings, this
can also be see clearly that the $\phi \bar T(\bar t) \sim 1/f$,
while in the s-channel, the couplings   $V\phi W $ ($ V=
A_H,~Z_L,~Z_H$) $\sim v'$, $v'=  xv^2/(4f)$, so they decrease with
increasing $f$. The exception, however, occurs in
FIG. (\ref{qqsw-1})(c)(d) and therefore FIG. (\ref{qqsw-2}).

 From FIG. (\ref{qqsw-1}),
we can also see that, when $x$ is small, such as $0.1$ which we have
chosen, in the most parameter space, the largest channel of the
processes $qq\to \phi W$ is the $b\bar b \to \phi W$, which is easy
to understand since, in FIG. (\ref{ppsw-lh}), the t-channel process (d)
is free of the center-of-mass suppression and the upraise via $x$
i.e, the $v'$ does not reveal itself. For larger $x$, however, the
situation changes. we can see from FIG. (\ref{qqsw-2}) that, except via
$\bbbar$ annihilation, the quark anti-quark processes are increasing
with the increasing $x$.

FIG. (\ref{qqsw-2}) is the cross sections of the quark anti-quark annihilations on the parameter $x$,
taking $x$ from $0$ to $0.9$, too.
In FIG. (\ref{qqsw-2}) we can see that with the increasing parameter $x=4fv'/v^2$,
the trends of the cross sections are different with those in FIG. \ref{ggsw-2}. When  $x> 0.2 $
the cross sections from the $u\bar u$ collision begin overwhelming that from $b\bar b$ at $14$ TeV,
 which is opposite to the above discussion. The reason is given in the following, the same as
 the comparison of $\phi W$ production via $gg$ fusion and $q\bar q$ annihilation.

Now, compared the FIG. (\ref{ggsw-2}) (c) with FIGs. (\ref{qqsw-2}), (\ref{ppsw-2}), we
can find that the $\phi W$ associated production from $gg$ fusion and $b\bar b$
annihilation
decreases with increasing $x$, while those from other quark anti-quark
annihilation ($u\bar u$, $d\bar d$, $s\bar s$ and $c\bar c$)
is opposite, which is understandable from their
different coupling forms. Since the  $V W_L \phi^+$ ($V = A_L,
~Z_L,~Z_H$) is  proportional to $v'$, while $\phi \bar t(T) b$ is in
proportion to $\frac {v}{f} -2s_+$, here $s_+ = 2v'/v$. In the
scalar-fermion couplings, actually, there is a competition between
the two terms  $v\over f$ and $2s_+$. When $f = 500$ GeV, $v/f\sim
0.5$, the $2s_+$, however, less than $0.5$ all the time if we
satisfy the requirement $v'<30$, i.e, $x<1$\cite{littlest}, so with
the increasing $v'$, the couplings $\phi \bar t(T) b$ is decreasing.

\def\figsubcap#1{\par\noindent\centering\footnotesize(#1)}
 \begin{figure}[t]%
 \begin{center}
   \parbox{6cm}{\epsfig{figure=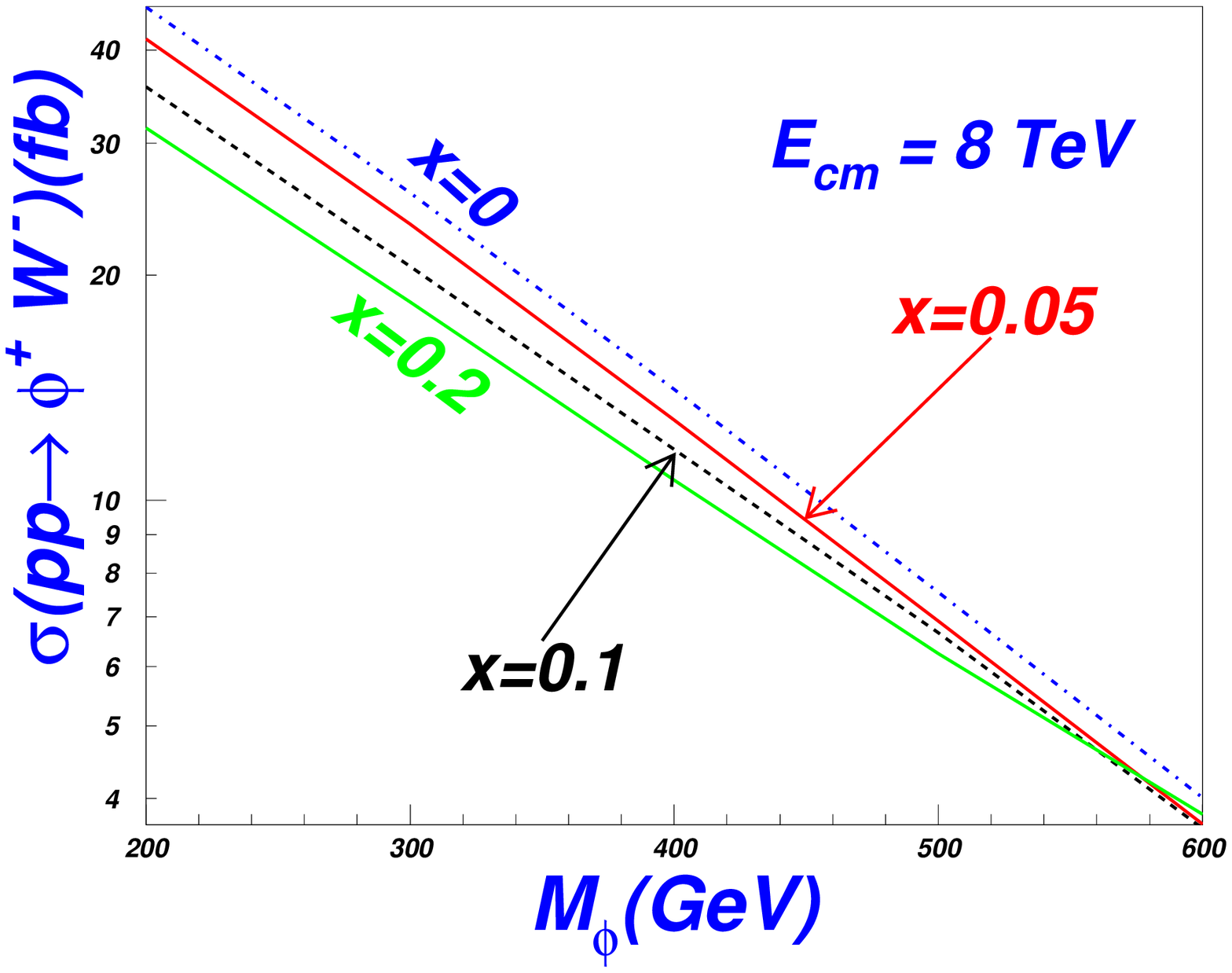,width=6cm}    \figsubcap{a}}
   \parbox{6cm}{\epsfig{figure=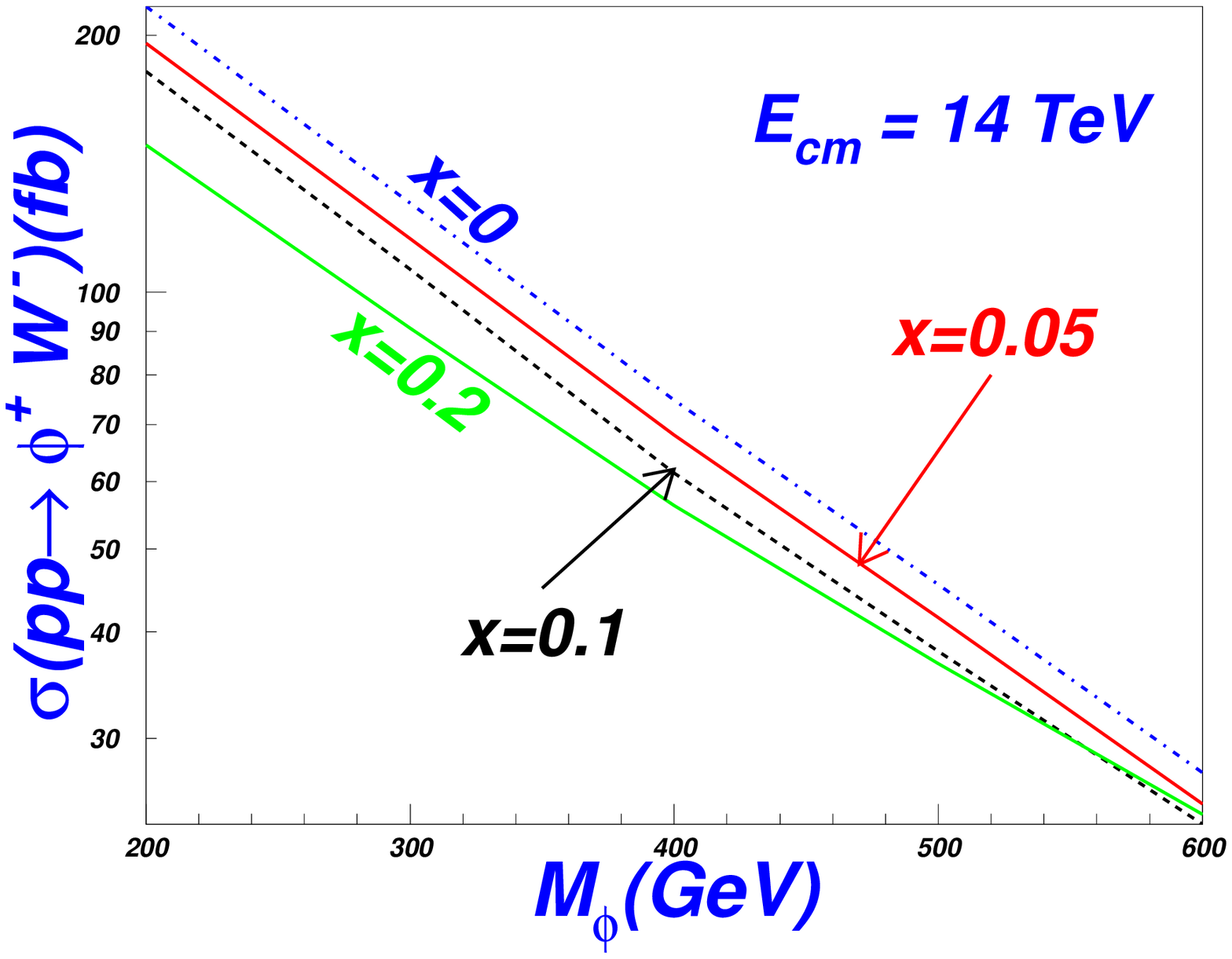,width=6cm}    \figsubcap{b}}
   \parbox{6cm}{\epsfig{figure=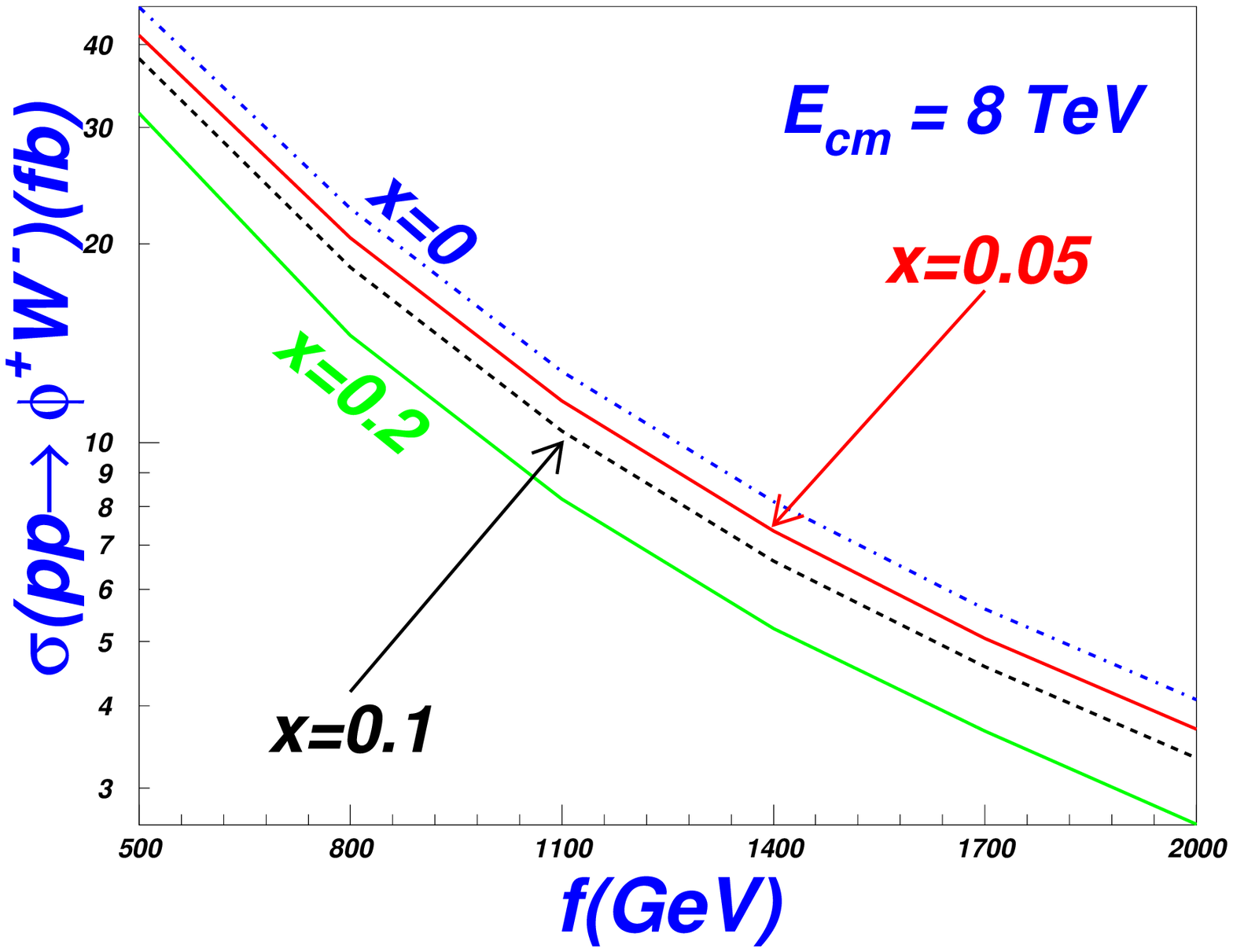,width=6cm}    \figsubcap{c}}
   \parbox{6cm}{\epsfig{figure=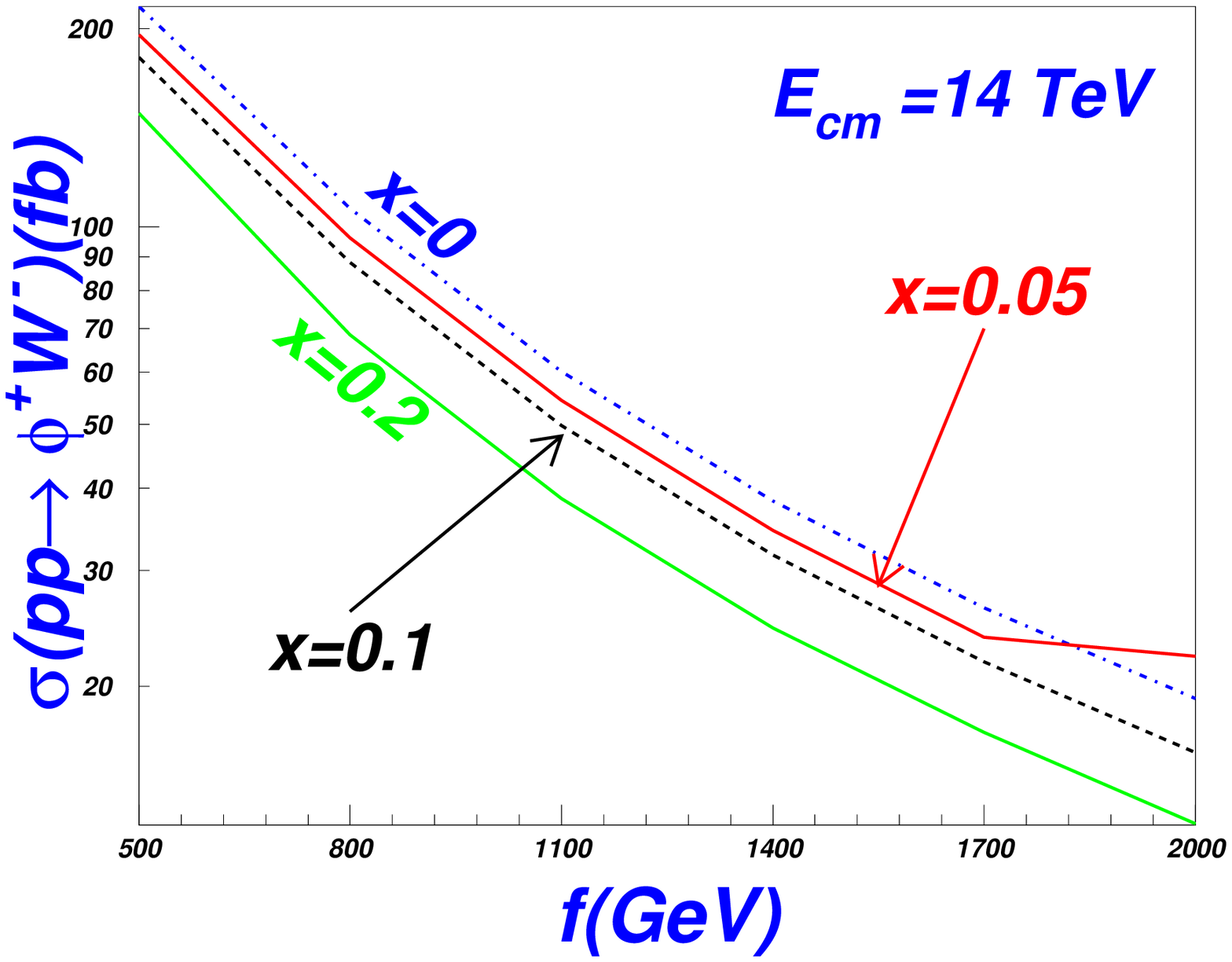,width=6cm}    \figsubcap{d}}
 \caption{ In LH models, the total cross section $\sigma$ of the processes $pp \to \phi^+W^-$
  as a function of the scalar mass $m_{\phi}$ with $E_{cm}=8$ TeV (a) and $14$ TeV (b) for $s=0.1$, $f=500$ GeV and
$x=~0, ~0.05,~0.1,~0.2$; The total cross section $\sigma$ of the
processes $pp \to \phi^+W^-$
  as a function of $f$ with the scalar mass $m_{\phi}=200$ GeV, $s=0.1$ and $E_{cm}=8$ TeV(c)
and $E_{cm}=14$ TeV (d) for $x=~0, ~0.05,~0.1,~0.2$.
\label{ppsw-1} }
 \end{center}
 \end{figure}
 \def\figsubcap#1{\par\noindent\centering\footnotesize(#1)}
 \begin{figure}[htb]%
 \begin{center}
  \parbox{7cm}{\epsfig{figure=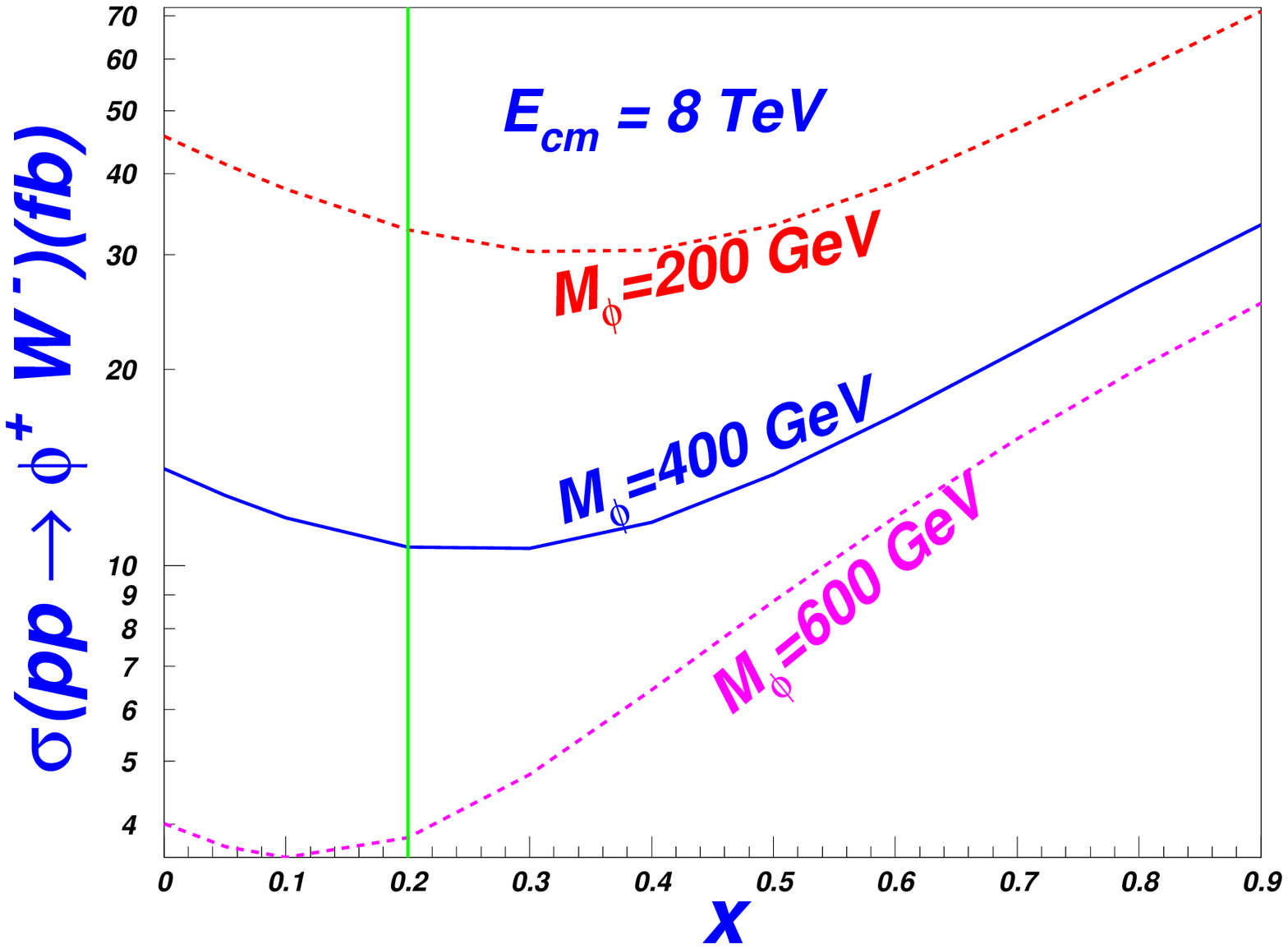,width=7cm}  \figsubcap{a}}
  \parbox{7cm}{\epsfig{figure=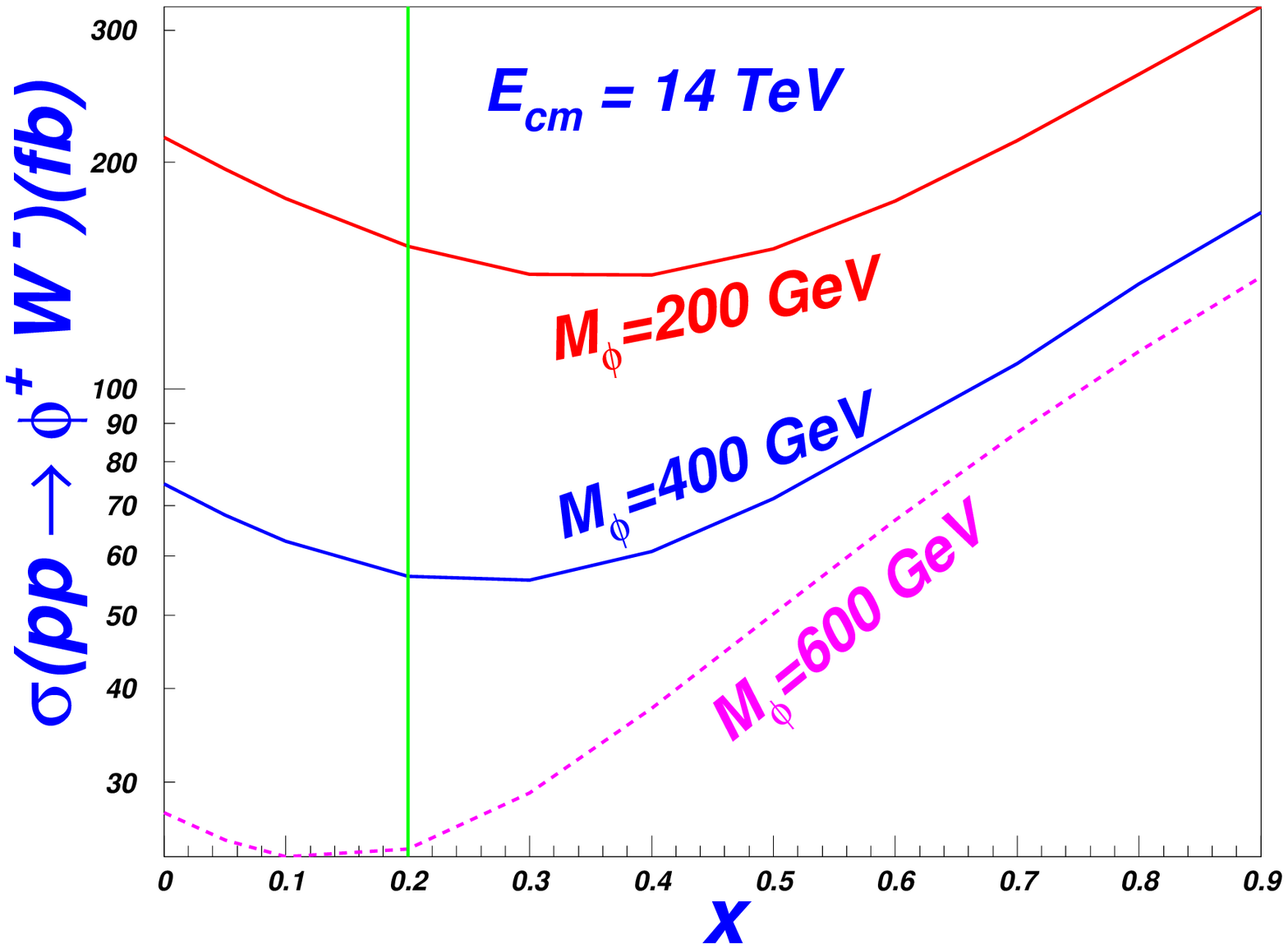,width=7cm}  \figsubcap{b}}
 \caption{ The total cross section $\sigma$ of the processes $pp \to \phi^+W^-$
  as a function of $x$ with the scalar mass $m_{\phi}=200,~400,~600$ GeV, $s=0.1$ and $E_{cm}=8$ TeV(a)
and $E_{cm}=14$ TeV(b). \label{ppsw-2} }
 \end{center}
 \end{figure}
\subsubsection{Total contribution of the gg and quark anti-quark annihilation }

In FIGs. (\ref{ppsw-1}), (\ref{ppsw-2}) we sum the contributions from all
the parton level processes. We can see from the figures that the
cross sections can arrive at tens of fb even when $E_{cm} =8$ TeV,
and when the center-of-mass rises to $14$ TeV, the production rates
will become more large, larger than $100$ fb in quite a large
parameter space. So in the discussion of reducing the backgrounds,
we will concentrate to the $14$ TeV center of mass.

With the increasing $x$, the s-channel contributions of the $q\bar q
$ annihilation contribute more and more large, so then the $gg$
fusion and $b\bar b$ collision, the t-channel dominant, are not the
largest any more, but instead, the $uu$ and $dd$ will control the
situation, which can be seen clearly in FIG. (\ref{ppsw-1}).

From FIG. (\ref{ppsw-1}) we can see that the production rates of the
$\phi^+W^-$ decrease when $m_\phi$ or $f$ goes up. Note that in
FIG. (\ref{ppsw-1}) (c)(d), with the increasing $f$, in the tail of the
curve for $pp\to \phi^+W^-$, $x=0.7$, the cross sections increase
when $f$ changes from $1500$ GeV to $1700$ GeV, which is
understandable, when $x$ is large, the contributions from the
s-channel surpass that from the t-channel, i.e, the $gg$ and $b\bar
b$ realization.

\section{ The $LRTH$ model and $\phi W $ production
at the $LHC$ }
\subsection{The $LRTH$ model and the relevant couplings}
To solve the little hierarchy problem \cite{littlehierarchy}, the left right twin higgs
models was proposed \cite{twinhiggsleftright,0611015-su}. In this models,
The higgses emerge again as pseudo-Goldstone bosons 
and the leading order of the the higgses masses is quadratically divergent.
One introduces an additional discrete symmetry so that the leading quadratically divergent terms respect the
global symmetry. With the cancellation of the quadratically divergent, then the higgs
masses posses logarithmically divergent contributions.

In such models, the global symmetry breaks from ${\rm U}(4) \times{\rm U}(4)$ to ${\rm
U}(3)\times{\rm U}(3)$, and gauge symmetry from
${\rm SU}(2)_L\times {\rm SU}(2)_R \times {\rm U}(1)_{B-L}$ to
${\rm SU}(2)_L \times {\rm U}(1)_Y$. 
Fourteen Goldstone bosons are generated, three of which are
eaten by the massive gauge bosons $Z_H$ and $W_H^{\pm}$, while the
rest of the Goldstone bosons contain the SM ${\rm SU}(2)_L$ higgs
doublet and extra higgses.

To cancel the leading quadratically divergence of the SM gauge bosons and the top quark contributions
to the higgs masses in the loop level, the new heavy gauge bosons and a vector top singlet pair are introduced.
Thus the hierarchy problem is solved. The new particles in the LRTH, both the gauge bosons and the vector top singlet,
have rich phenomenology at the LHC and people are interested in them.

The two higgs fields, $H$ and $\hat{H}$ acquire two non-zero VEVs
which break the $U(4)\times U(4)$ to $U(3)\times U(3)$ and yields 14 Goldstone bosons, six of which
are eaten by the massive gauge bosons. Finally, there are one neutral pseudoscalar
$\phi^0$, a pair of charged scalar $\phi^\pm$, the SM
physical higgs $h$, which representation of ($\phi^+,\phi^0$) is (1,2,1) 
in the gauge group
$SU(3)_L\times SU(3)_R\times U(1)_{B-L}$.  A ${\rm SU}(2)_L$ doublet
$\hat{h}=(\hat{h}_1^+, \hat{h}_2^0)$ are also left in the higgs spectrum.

The quantum numbers of gauge bosons of $W^\pm$ and $W_H^\pm$ are (3,1,0) and (1,3,0).
After the symmetry breaking, the six physically massive gauge bosons are four charged and two
neutral ones: $W^{\pm}$, $W^{\pm}_H$, $Z$ and $Z_H$. $W$
and $Z$ are the usual massive gauge bosons in the SM and $W_H$, and
$Z_H$ are three additional new massive gauge bosons with masses of
TeV.

The Lagrangian concerning of the new particles can be written as
\begin{eqnarray}\label{lagrangian}
    {\cal L} = {\cal L}_H+{\cal L}_G+{\cal L}_f+{\cal L}_Y+{\cal L}_{one-loop}
+{\cal L}_{\mu}.
 \end{eqnarray}
The various terms in Eq.~(\ref{lagrangian}) are covariant kinetic terms
for higgses, gauge bosons and
fermions, Yukawa interactions, one-loop Coleman-Weinberg (CW)
potential~\cite{CW,0611015-su} for higgses and soft symmetry breaking $\mu$
terms. The explicit expression can be found in Ref. \cite{0611015-su} and we here not list repeatedly.

Based on the Lagrangian given in Ref. \cite{0611015-su},
 we have, 
the couplings with fermions involved, which are concerned of our calculation \cite{0611015-su},
\begin{table}[h]
\begin{tabular}{|c|c||c|c|}
\hline
particles & vertices & particles & vertices \\
\hline $W_{+\mu} \bar t b$ &  $ ~~e \gamma_{\mu} C_LP_L/( \sqrt{2} s_w) $ &
$W^{+\mu} \bar T b$ &    $ ~~e \gamma_{\mu} S_LP_L/( \sqrt{2} s_w)  $ \\
\hline $H \bar t t$ &$          -e m_t C_L C_R/(2 m_{W} s_w) $ &
$H \bar T T$ & $      -y (S_R S_L-C_L C_R x)/\sqrt{2}    $ \\
\hline $\Phi^0 \bar t t$ & $   -i yS_R S_L \gamma_5 /\sqrt{2}  $ &
$\Phi^0 \bar T T$ &$   -i yC_L C_R\gamma_5/\sqrt{2}    $  \\
\hline $\Phi^+ \bar t b$ & $  -i (S_R m_b  P_L-y  S_L f P_R)/ f        $ &
$\Phi^+ \bar T b$ & $   ~~i (C_R m_b  P_L-y C_L f P_R)/f          $\\
\hline
\end{tabular}
\caption{The three-point couplings of  the charged gauge
boson-fermion-fermion and those of the scalar-fermion-fermion in the LRTH models.
The chirality projection operators are $P_{R, L}=(1\pm\gamma_5)/2$.
\label{lrth-couplings}  }
\end{table}

As for the coupling between the boson and the scalars,  we find that
they all vanish, if we parameterize the scalars in the Goldstone
bosons fields as\cite{0611015-su},
\begin{eqnarray}
\label{unitary}
    \begin{array}{ll}
  N~ \to  \frac{\sqrt{2}\hat{f}}{F(\cos x+2\frac{\sin
x}{x})}\phi^0, &
~\hat{N}~ \to  -\frac{\sqrt{2}f\cos x}{3F}\phi^0, \\
  h_1\to  0, &
~h_2 \to
\frac{v+h}{\sqrt{2}}-i\frac{x\hat{f}}{\sqrt{2}F(\cos x+2\frac{\sin
x}{x})}\phi^0, \\
  C~ \to  -\frac{x\hat{f}}{F\sin x} \phi^+, &
~\hat{C}~ \to  \frac{{f\cos x}}{F} \phi^+. \\
\end{array}
\end{eqnarray}
where the $N, ~ \hat{N},~ h_1,~h_2,C,~\hat{C}$ are in the Goldstone
bosons fields,
\begin{equation}
\label{GBrep}
    H= i\frac{\sin\sqrt{\chi}}{\sqrt{\chi}}e^{i\frac{N}{2f}}\left(%
\begin{array}{c}
  h_1 \\ h_2\\
  C\\
  N-if\sqrt{\chi}\cot\sqrt{\chi}\\
\end{array}%
\right), \ \ \hat{H} =
i\frac{\sin\sqrt{\hat{\chi}}}{\sqrt{\hat{\chi}}}
e^{i\frac{\hat{N}}{2\hat{f}}}\left(%
\begin{array}{c}
  \hat{h}_1 \\ \hat{h}_2\\
  \hat{C}\\
  \hat{N}-i\hat{f}\sqrt{\hat{\chi}}\cot\sqrt{\hat{\chi}}\\
\end{array}%
\right).
\end{equation}

By this parameterization, the requirement of vanishing gauge-higgs
mixing terms can be satisfied, i.e, in this redefinition of the
higgs fields, the couplings $WZ\phi^+$, $W\gamma\phi^+$,
$WZ_H\phi^+$, $W\gamma_H\phi^+$, $W\phi^0\phi^+$, and $Wh\phi^+$ are
zero, which has been
verified by our written calculation. This is quite different with that in the littlest higgs models.


\subsection{ LRTH $\phi W$ production  at the LHC and the numerical results }
Due to  the missing  of the gauge-higgs mixing terms, the associated
production of the charged scalar $\phi^+$ and  the charged gauge
boson $W$ is different with that in the little higgs models. In
FIG. (\ref{ppsw-lh}), the figures (a) and (e) will not occur in the
LRTH models since they contain the gauge-higgs mixing couplings,
while the others are kept and they are the realization of the $\phi
W$ production in the LRTH models.

When discussing the numerical results of the  processes, just as
the discussions in LH models, we also, firstly, investigate the
contributions from every single parton channel, i.e,  the $gg$
fusion and the $q\bar q$ annihilation processes, respectively, and
then sum them for the total contributions.

\subsubsection{gg fusion in the LRTH models}
 \def\figsubcap#1{\par\noindent\centering\footnotesize(#1)}
 \begin{figure}[t]%
 \begin{center}
   \parbox{6cm}{\epsfig{figure=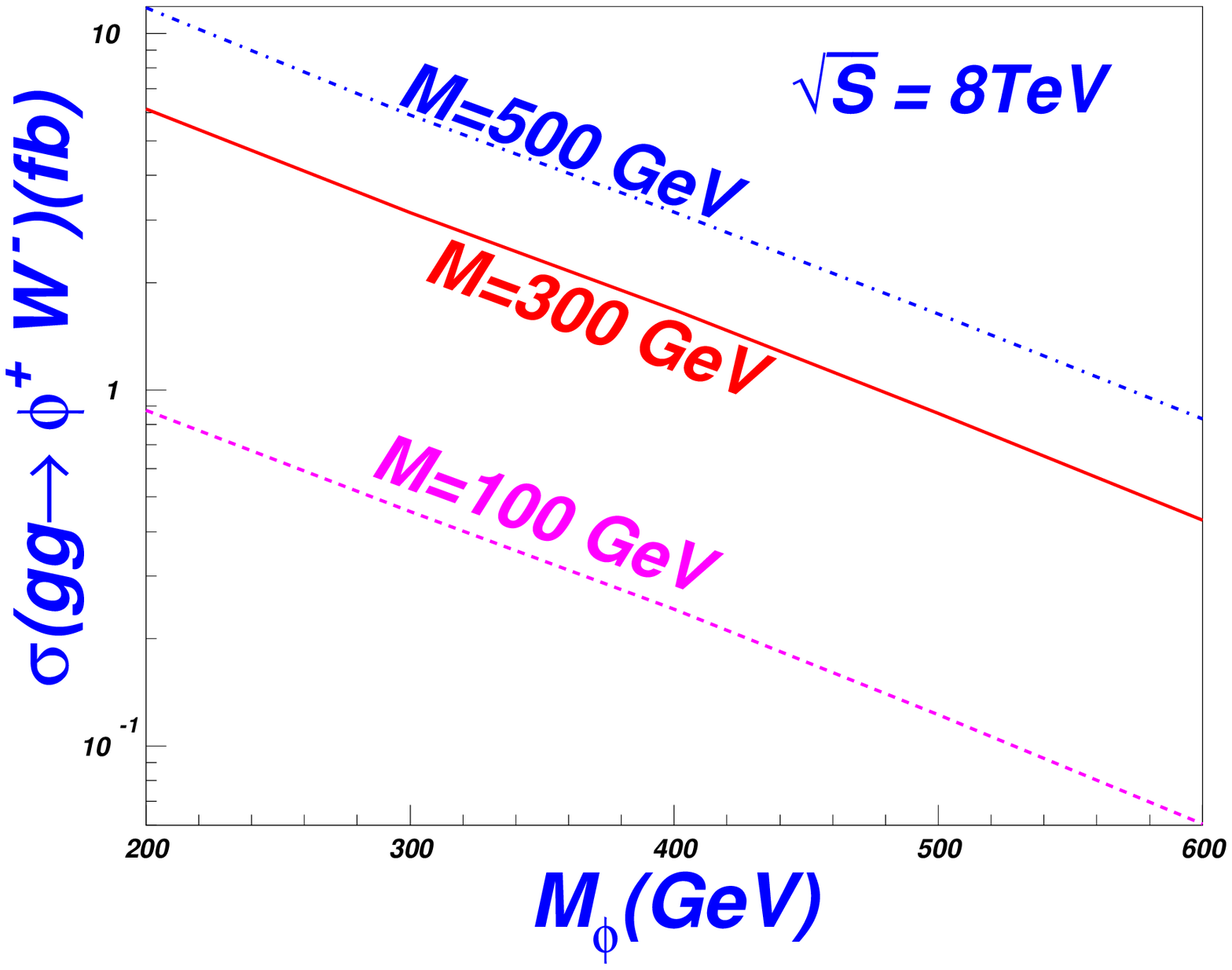,width=6cm}  \figsubcap{a}}
   \parbox{6cm}{\epsfig{figure=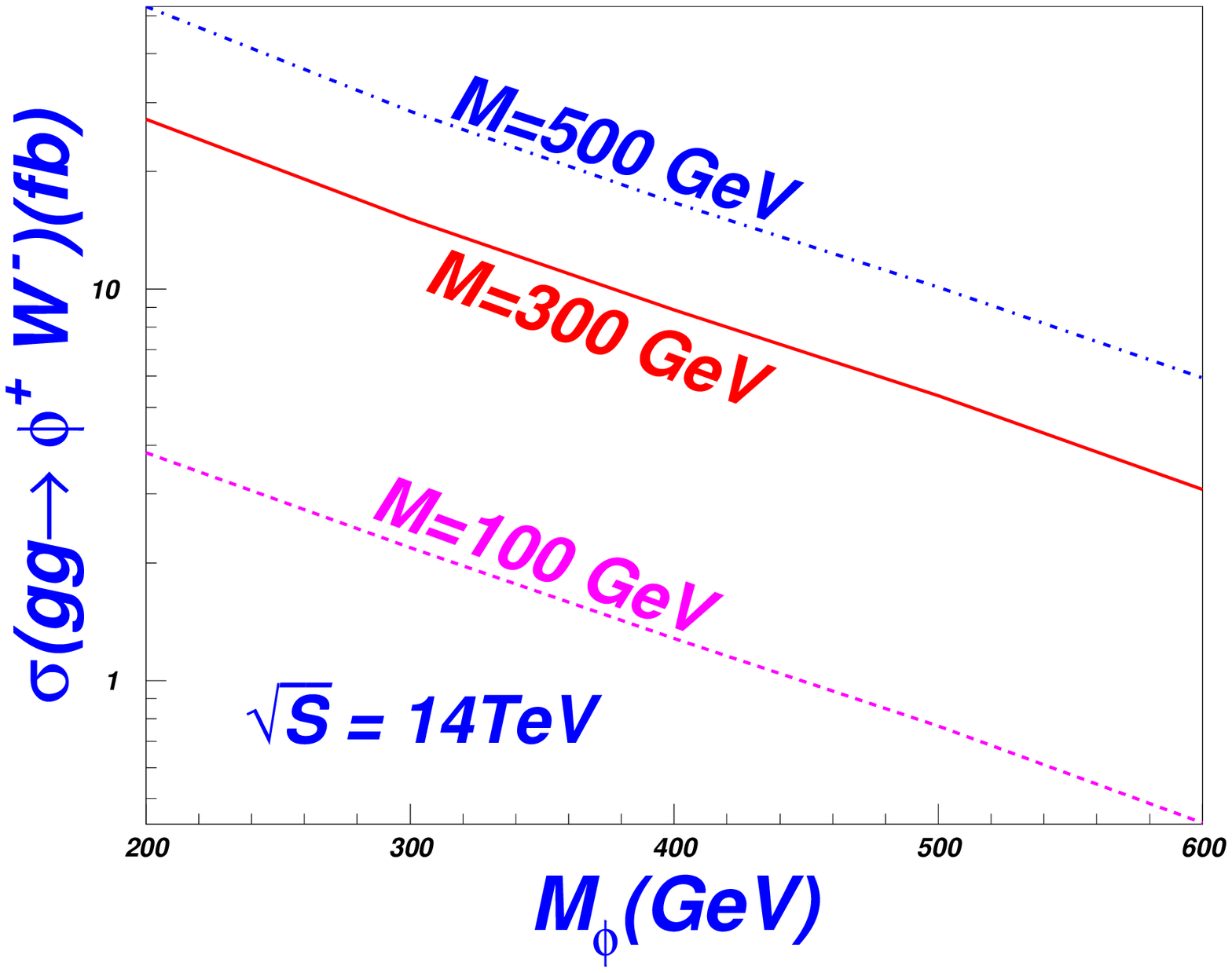,width=6cm} \figsubcap{b}}
   \parbox{6cm}{\epsfig{figure=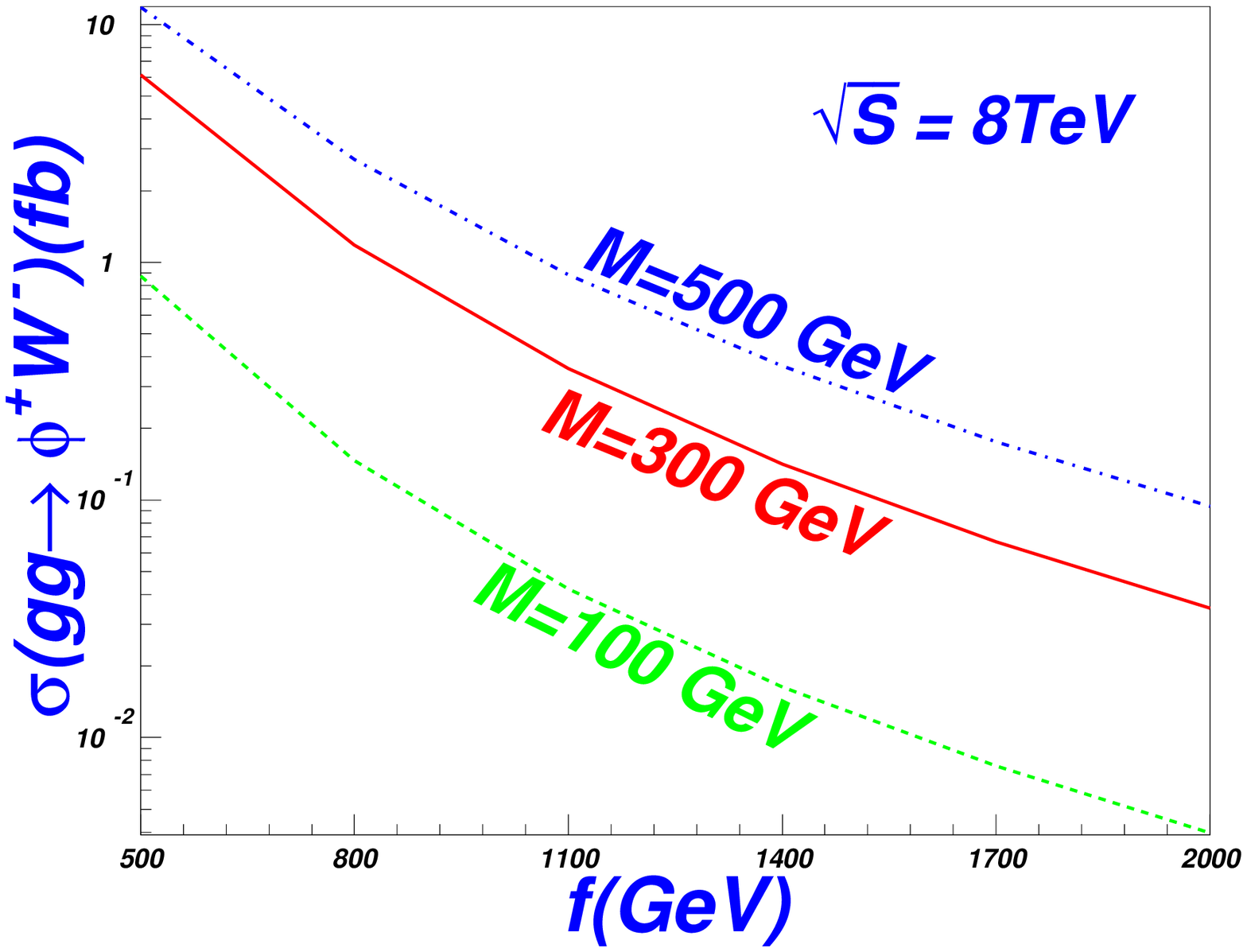,width=6cm}  \figsubcap{c}}
   \parbox{6cm}{\epsfig{figure=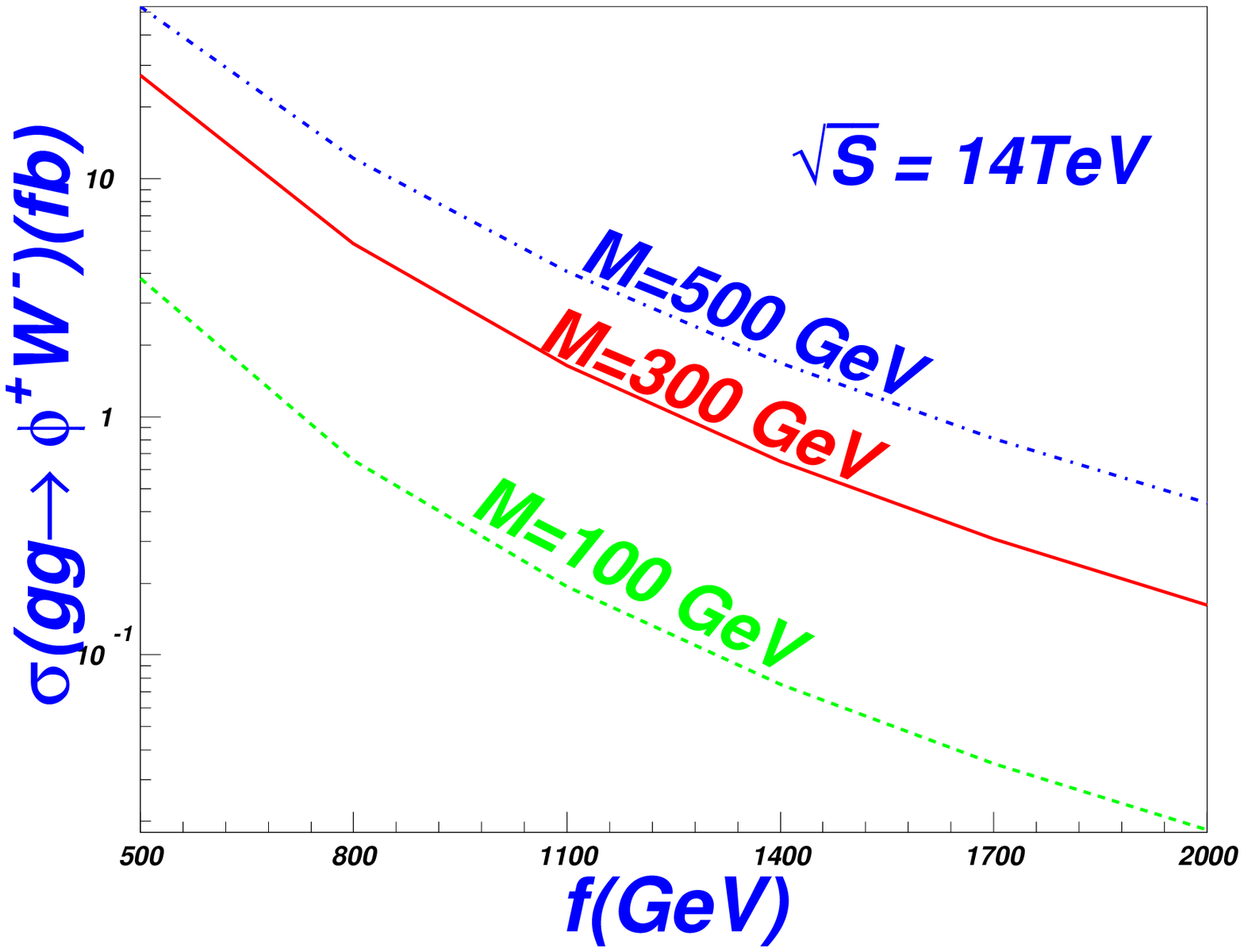,width=6cm}  \figsubcap{d}}
 \caption{ In LRTH, the cross section $\sigma$ of the processes $gg \to \phi^+W^-$
  as a function of the scalar mass $m_{\phi}$ or $f$ with $E_{cm}=8$ TeV
and $E_{cm}=14$ TeV for $M=~100,~300,~500$ GeV.
\label{ggsw-3} }
 \end{center}
 \end{figure}

Different with that of the LH models, the $\phi W$ associated
production are carried out only by the box diagrams from $gg$ fusion
and t-channel contribution via  the quark anti-quark annihilation,
just shown as FIG. (\ref{ppsw-lh}) (b)(c) and (d), and the s-channels in
FIG. (\ref{ppsw-lh}) (a) (e) are missing.

The production cross sections of the $\phi^+ W^-$ of the $gg$ fusion
are plotted in FIG. (\ref{ggsw-3}) with $M=100,~300,~500$ GeV for
$E_{cm}=8,~14$ TeV and for $f=500$ GeV, as functions of the
scalar mass $m_\phi$, assuming the charged and neutral scalar mass
degenerate, $m_{\phi^\pm} =m_{\phi^0}=m_{\phi^p}$. From
FIG. (\ref{ggsw-3}), we can see the cross section of this process is
less than  $50$  fb in most of the parameter space, even for
a larger certer-of-mass energy, i.e, at $14$ TeV with $M=500$ GeV.
We can also see that, as expected, the production rate decreases with
the increasing scalar mass since the phase space are suppressed by the mass.

FIG. (\ref{ggsw-3}) show the different dependence of the cross
sections on the parameter $M$, with $M=100,~300,~500$ GeV.
The results change with the varying values of $M$ and when $M$ is large, such as

When $M$ is very small,such as $\lesssim$
1 GeV, the collider phenomenology of the $\phi^+ W^-$ will very small,
which can be seen clearly via the two group couplings that realize the $\phi^+W^-$
associated production. The $\phi^+\bar t b$ and $W_\mu^- t\bar b$ couplings, for example,  are
$(S_R m_b  P_L-y  S_L f P_R)/ f $ and $ \gamma_{\mu} C_LP_L/( \sqrt{2} s_w) $, respectively,
with $S_L,S_R \sim M/M_T$ and $C_L = \sqrt{1-S_L^2}$. So when $M$ is small,  $S_L,S_R$ will
become small too. When $M =0$, $S_L,S_R$ also change into 0. So if $M$ is too small, the signal
 will be very small. In the limit
case, when $M=0$, the light top will not mix with the heavy top, so the couplings
$\phi^+\bar t b$ disappear, and the contribution are only from the heavy
top coupling to the scalar. But the light charged scalar is oppsite, which, mainly
coupled to the light top, the heavy top couplings $W^+\bar T b$, proportional to $S_L\sim M$,
disappear. So the cross section will drop down to zero when $M$ is in its limit $M=0$.

We can also see from FIG. (\ref{ggsw-3}) that the process $gg\to \phi^+W^-$
is dependent strongly on the parameter $f$, which is understandable
since, the most couplings in the LRTH models, such as $\phi^+ t\bar
b$),$\phi^+ T\bar b$, etc, are tightly connected with the parameter
$f$. The cross sections may be larger unless $f$ is not too high. The
rates of the $\phi^+W^-$ production for $\sqrt s = 14 $ TeV and
 $m_\phi =200$ GeV, for example, are $52$ fb and $7$ fb, for $f=500$ GeV and
$f= 1000$ GeV, respectively.

\subsubsection{$b\bar b$ annihilation in the LRTH models}
\def\figsubcap#1{\par\noindent\centering\footnotesize(#1)}
 \begin{figure}[t]%
 \begin{center}
  \parbox{6cm}{\epsfig{figure=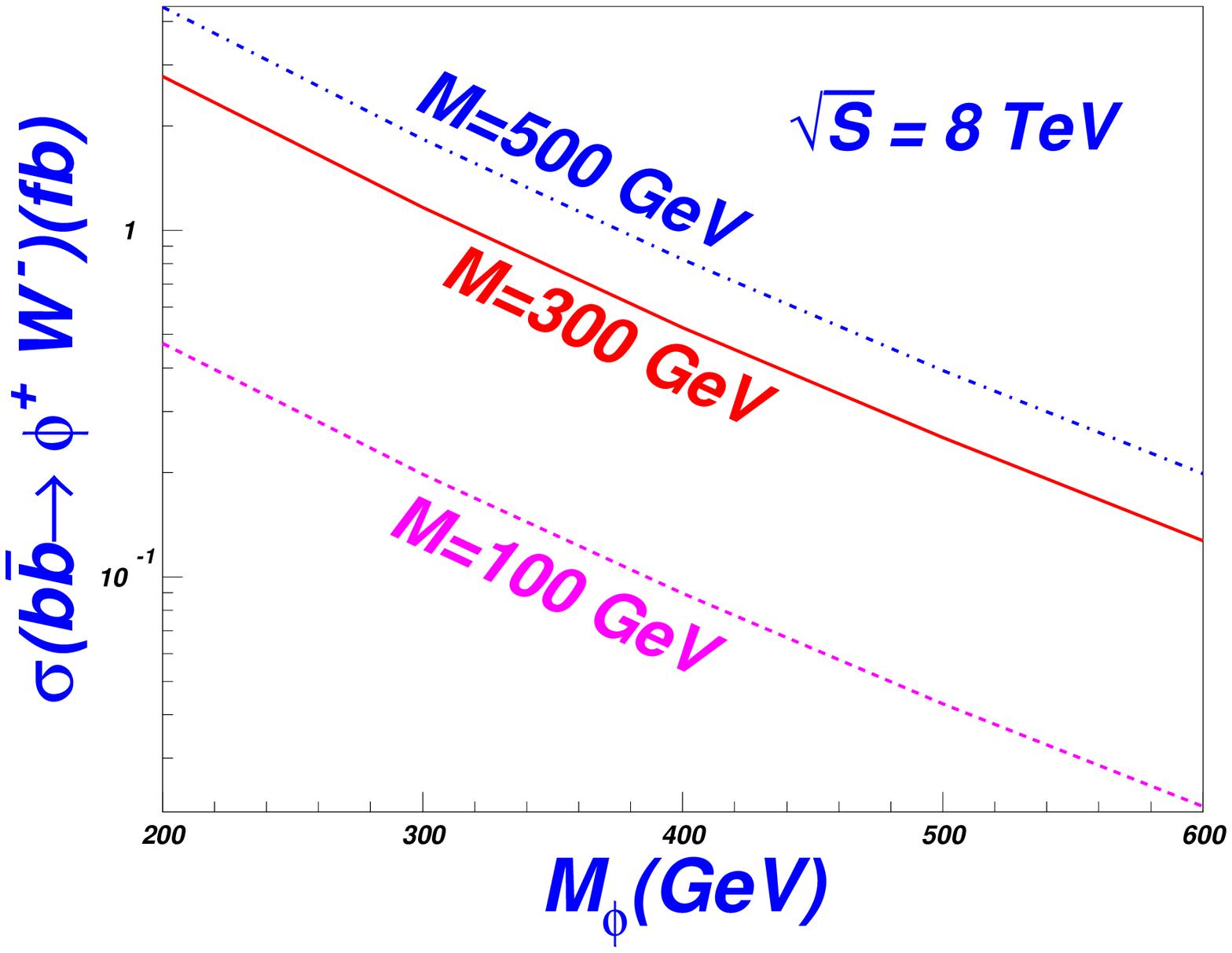,width=6cm}  }  
  \parbox{6cm}{\epsfig{figure=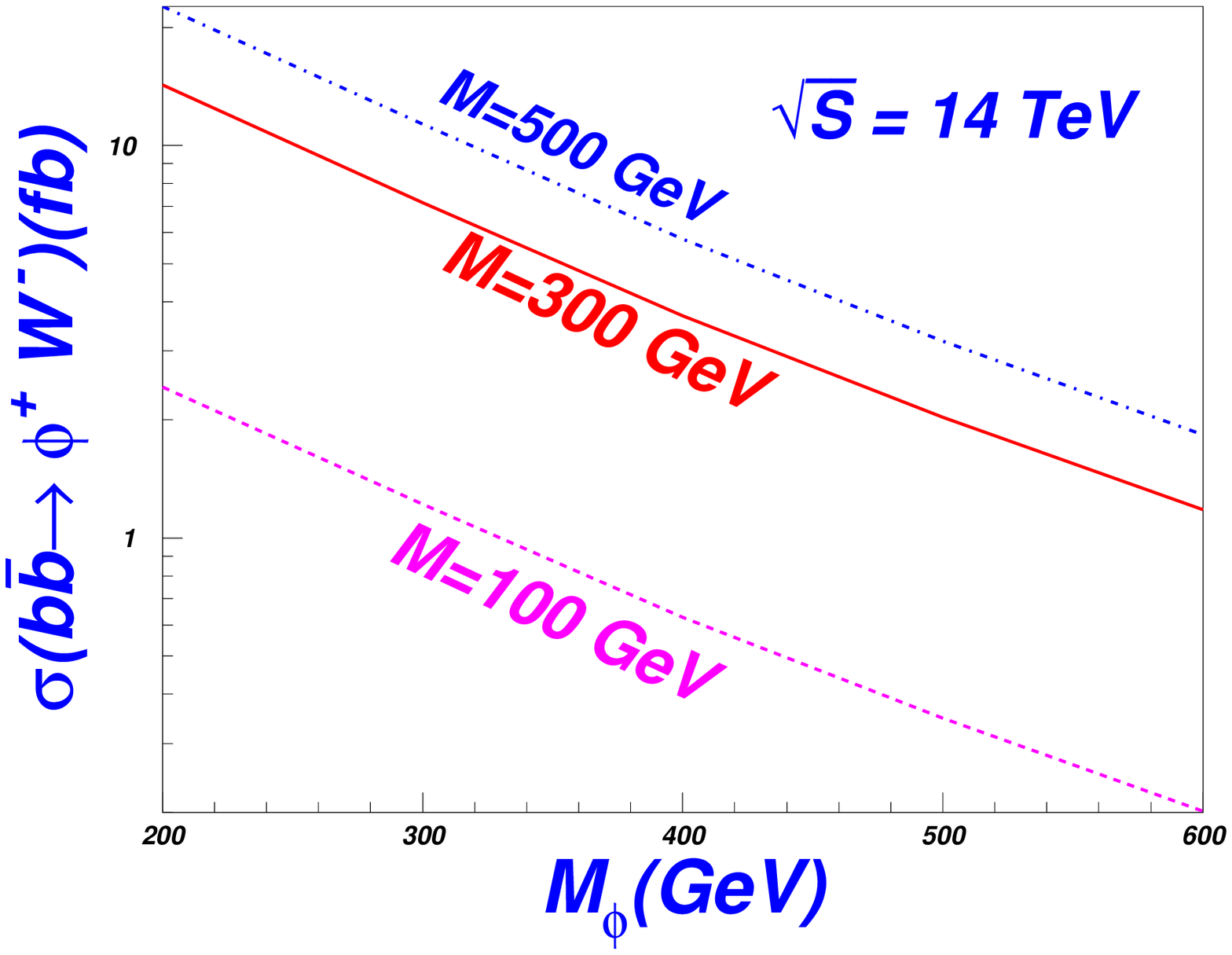,width=6cm} }  
  \parbox{6cm}{\epsfig{figure=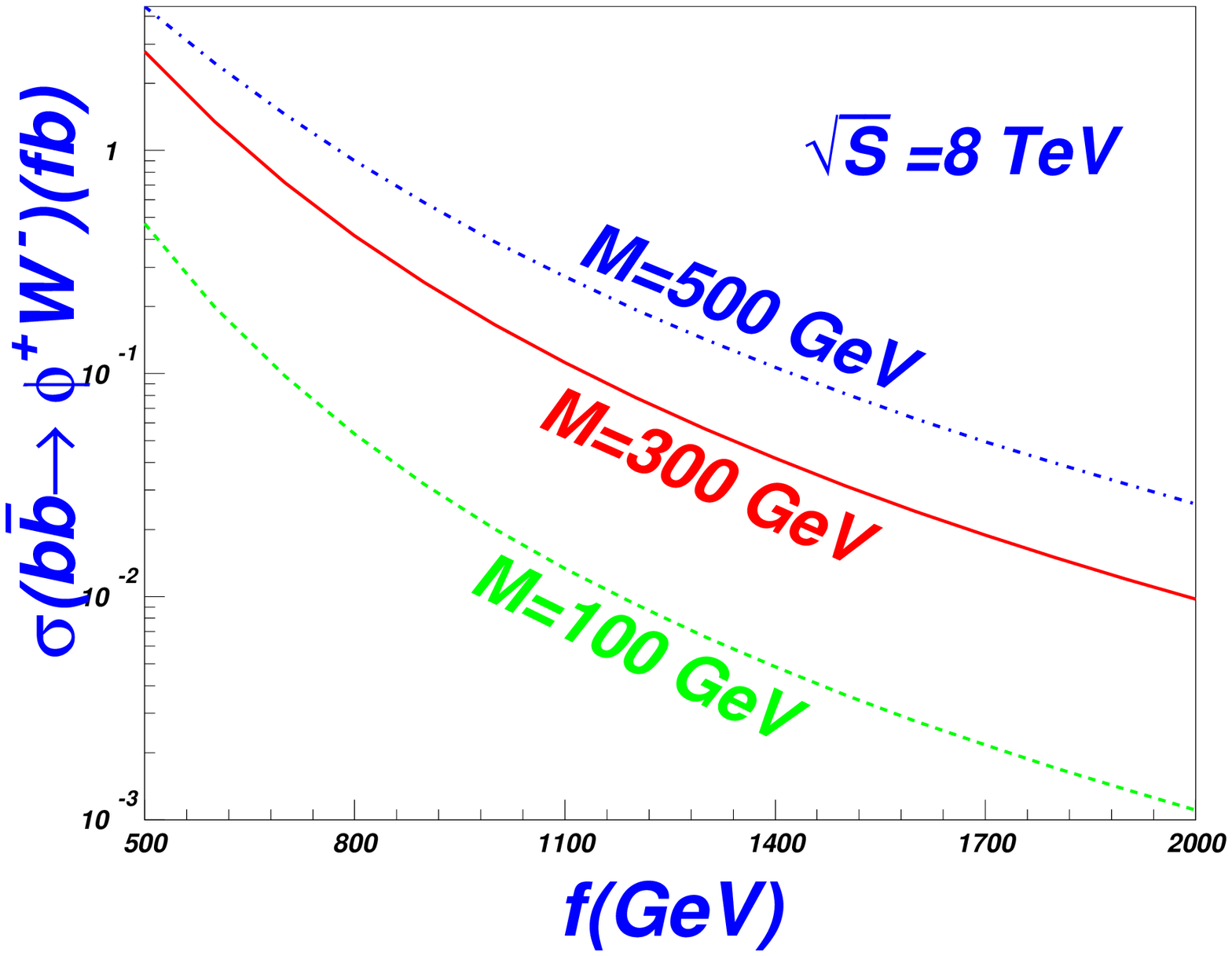,width=6cm}  }  
  \parbox{6cm}{\epsfig{figure=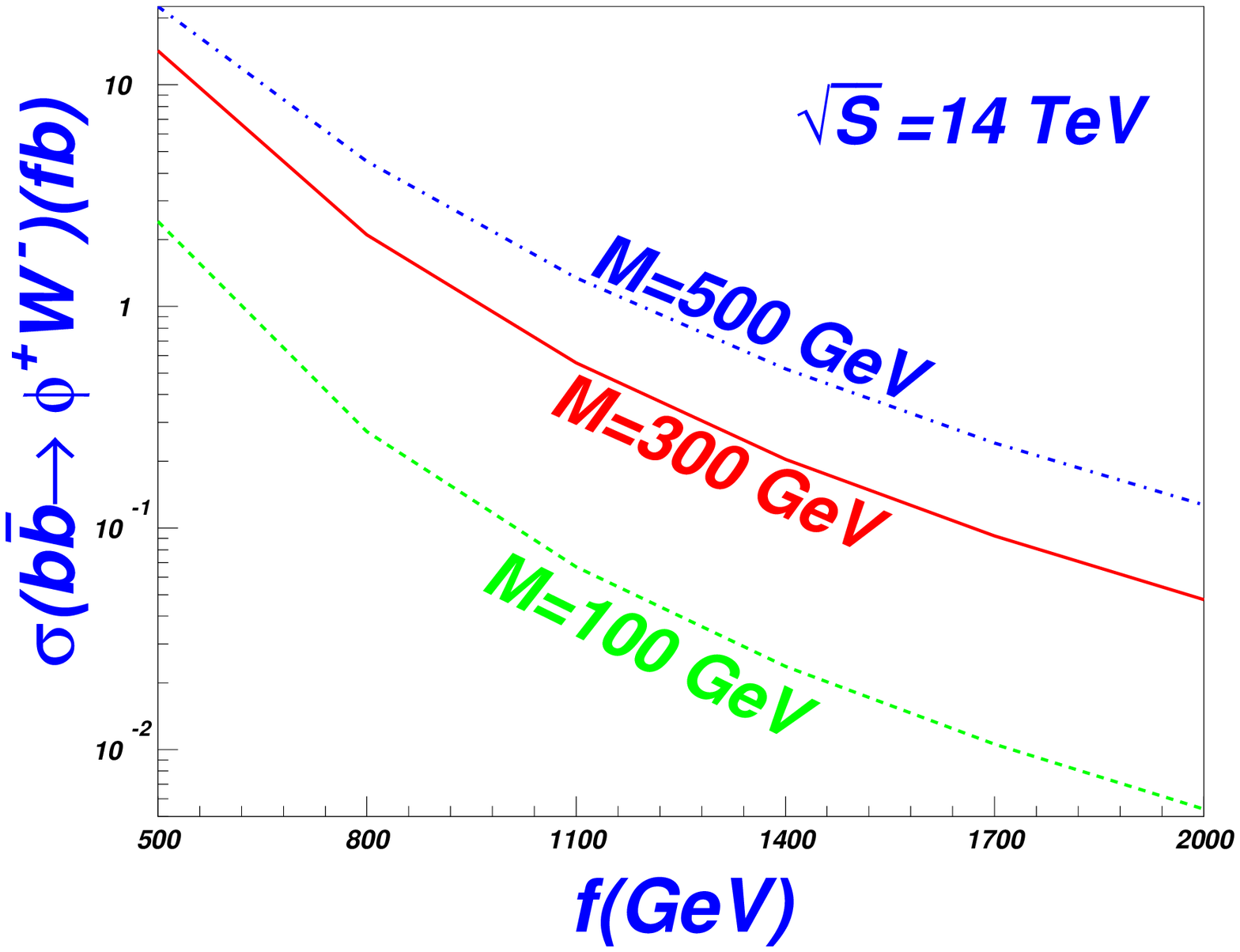,width=6cm} }  
 \caption{ In LRTH, the cross section $\sigma$ of the processes $q\bar q \to \phi^+W^-$
  as a function of $m_\phi$ or $f$ for $M=~150,~250,~500$ GeV with the scalar mass
$m_{\phi}=200$ GeV and $E_{cm}=8$ TeV and $E_{cm}=14$ TeV . \label{qqsw-3} }
 \end{center}
 \end{figure}

Unlike that in the LH models, in LRTH, the $\phi W$ production via
quark anti-quark annihilation are realized only by the t-channel
parton level $b\bar b $ $ \to  \phi W$, which is because we have
expected the higgs-gauge coupling vanishing, so the s-channel
processes are missing, and only the t-channel processes proceeding by
the $t-b$ and $T-b$ mixings survive.

Due to the small parton distribution functions, the $b\bar b $
realization of the $\phi W$ production, which is tree-level, is
not quite large, can arrive at about $20$ fb, a little smaller than  that
of the $gg$ fusion in the loop level realization.

At the same time, we can see that the
process $b\bar b  \to \phi W$ depends largely on the parameter $M$ and
$f$, and if $f$ goes up,   the production  rate of this process will decreases,
but for parameter $M$, the cross sections will increase with the increasing parameter $M$,
which can be also seen in FIG. (\ref{qqsw-3}).

\subsubsection{Total Contribution of the gg and quark anti-quark Annihilation }

\def\figsubcap#1{\par\noindent\centering\footnotesize(#1)}
 \begin{figure}[t]%
 \begin{center}
  \parbox{6cm}{\epsfig{figure=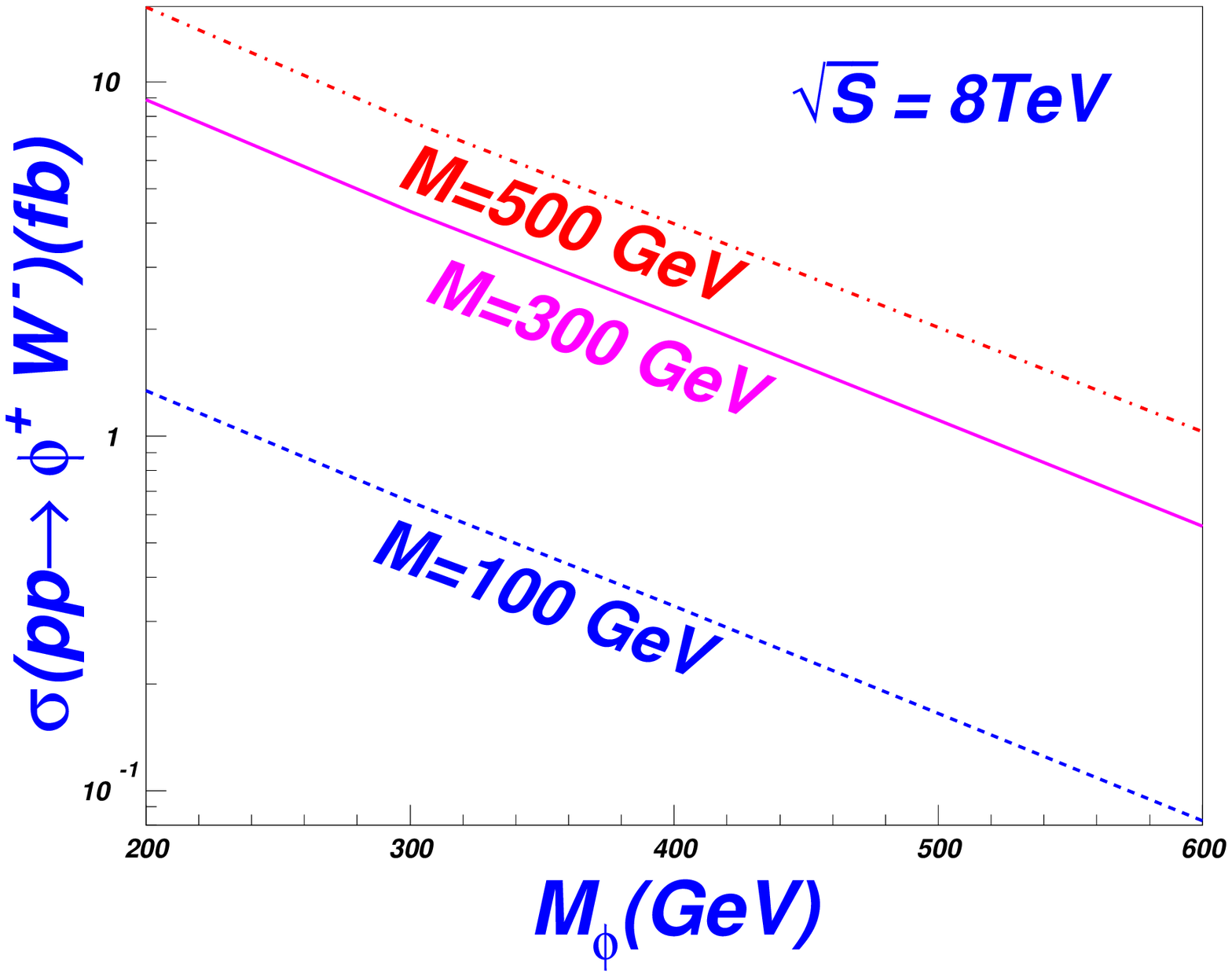,width=6cm}   \figsubcap{a}}
  \parbox{6cm}{\epsfig{figure=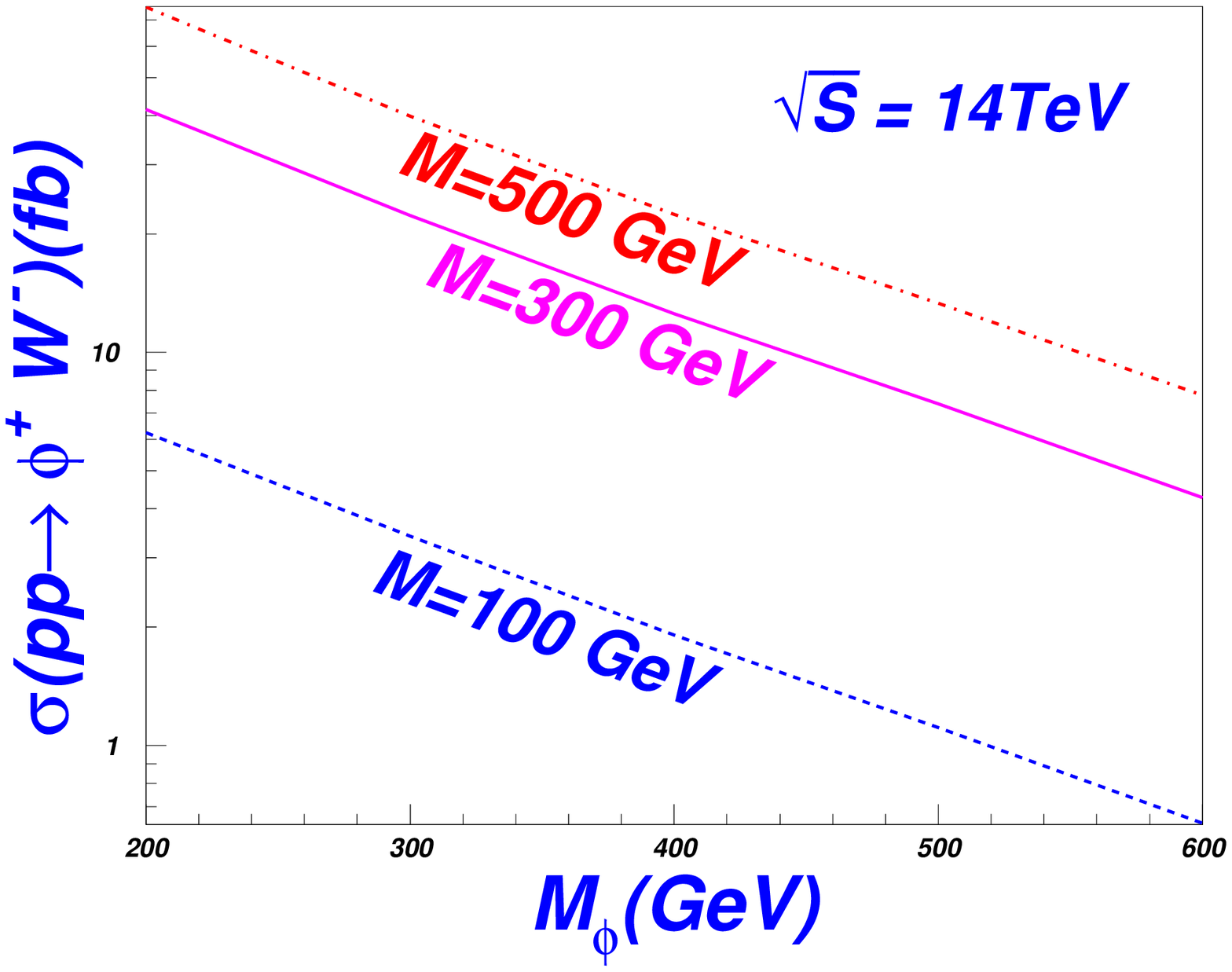,width=6cm}   \figsubcap{b}}
  \parbox{6cm}{\epsfig{figure=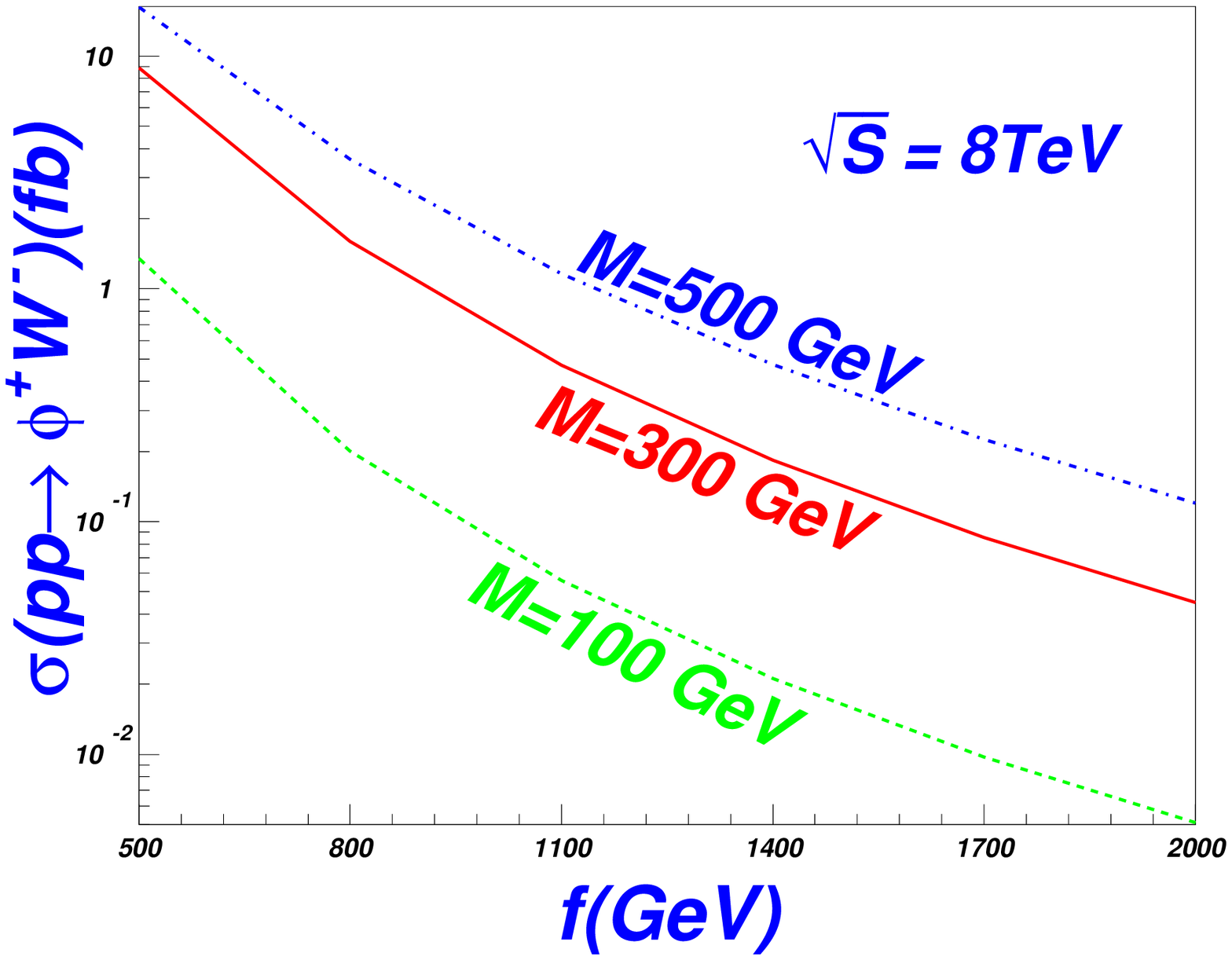,width=6cm}  \figsubcap{c}}
  \parbox{6cm}{\epsfig{figure=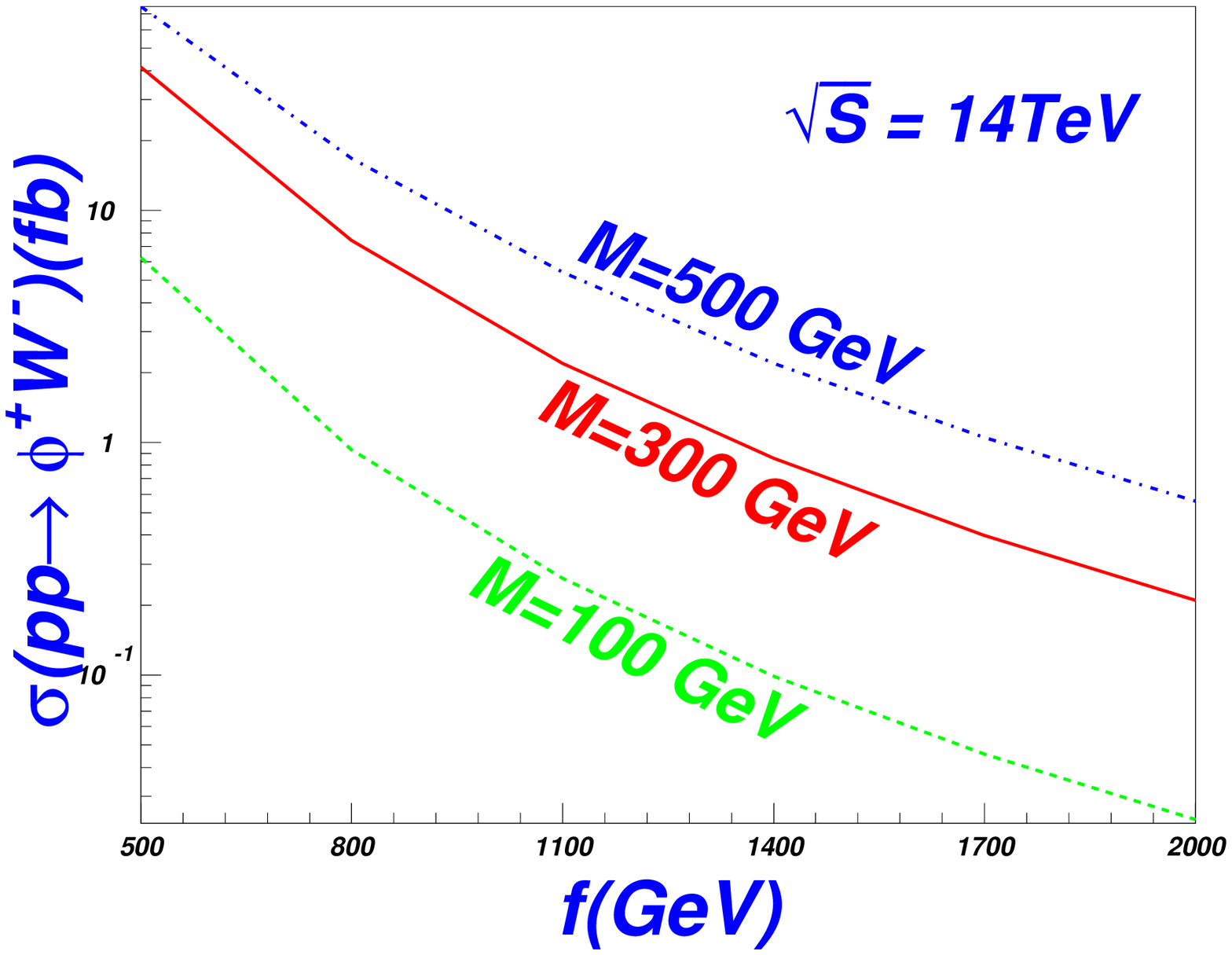,width=6cm}    \figsubcap{d}}
 \caption{ In LRTH,  the total cross section $\sigma$ of the processes $\phi^+W^-$
associted production from $gg$ fusion and $b\bar b$ collision
  as a function of the scalar mass $m_{\phi}$ or $f$ with $E_{cm}=8$ TeV
and $E_{cm}=14$ TeV for $f=500, ~1000$ GeV for different $M$ ($M=~100,~300,~500$ GeV).
\label{ppsw-3} }
 \end{center}
 \end{figure}
In LRTH models, we sum all the contribution, from $gg$ fusion and $b\bar b$ annihilation
for the $\phi^+ W^-$ associated production in FIG. (\ref{ppsw-3}), and from which, we can
 see that the cross section can arrive at tens of fb, dependent on the parameter
$f$, $M$ and the scalar mass in a certain center-of-mass $E_{cm}$. But in quite a large parameter space,
the cross sections are less than $10$ fb. Normally, at the LHC, this will not interest us, so we will
discuss little, also in the following section.

\section{Backgrounds and Detections}
From the data above, we can see that, at $E_{cm}=8$ TeV, no matter LH or LRTH models, the cross
section of the the charged higgs associated with a $W$ boson
production is quite small, even with a little scalar mass, such as
$200$ GeV, supposing the luminosity to be $10~fb^{-1}$. It is
easier, however,  for the charged higgs boson to be observed at
$E_{cm}=14$ TeV. Therefore from now on we focus on investigating
the charged higgs associated with a $W$ boson in the following
processes at the $14$ TeV. The following signatures can be considered\cite{11120086}:
\begin{eqnarray}
&pp&\to W^{-}H^{+}\to W^{-}t\bar{b}\to l^{-}\nu b\bar{b}jj, \nonumber \\
&pp&\to W^{+}H^{-}\to W^{+}\bar{t}b\to l^{+}\nu b\bar{b}jj
\label{pro:decay}
\end{eqnarray}
at $E_{cm}=14$ TeV with $200\leq m_\phi \leq 600$ GeV.

For the processes above with final state $ l+\met+b\bar{b}jj$, the
dominant SM backgrounds are $t\bar{t}$, $t\bar{t}W$, $t\bar{t}Z$,
$WZjj$, $WWjj$ and $Wjjjj$, which are discussed in
Ref. \cite{11120086}. In the signature of the $H^{\pm}W^\mp$ production processes,
the charged higgs decays to four jets and top quark decay to three of them.
 Thus to make clear of the signal, one can make the requirements like these: 1)
 the invariant mass of final four jets is around the charged
higgs mass, and  2) three jets of the four reconstruct into top quark mass.
To suppress the $t\bar{t}$ final state, the dominant channels of the backgrounds, one can
construct the second top quark.
 The final results given in Ref. \cite{11120086} show that after all
cuts, the signal process is only $1$ fb left when $m_{H^{\pm}}=500$ GeV,
and the backgrounds are becoming negligibly small.
 Ref. \cite{11120086} also points out
in Table I and
II, with the increasing charged scalar mass, the backgrounds become
smaller and easier to be suppressed, so it seems that the larger the
charged scalar mass is, the easier to detect the $WH$ production at
the LHC, though the cross section of the signals will also be
smaller.

 From Ref. \cite{11120086} Table II, we can see that if the scalar
 mass is $400$ GeV, the $S/\sqrt{B}$ can reach $3.42$, and with the
 increasing $m_S$ (scalar mass), the $S/\sqrt{B}$ gets larger, so we will focus
 the scalar mass at $400$ GeV and larger. From Table I of Ref. \cite{11120086}, we can see
 if $m_S = 400$ GeV, when the cross section arrive at $49.7$ fb, the
 $ S/\sqrt{B}$ will be larger than $3$.

Table \ref{cross-section} give the optimum value of the $\phi^+W^-$ production in the
LH and the LRTH models at the $14$ TeV when $m_\phi=400$ GeV. The parameters are set as:
 1) In the LH models, $s=0.1$, $s'=0.5$, $f=500,~1000$ GeV.
 2)In the LRTH models, the involved parameters are  $Y=1$, $f=500,~1000$ GeV.
\begin{table}[htb]
\begin{centering}
\begin{tabular}{|c|c|c|c|c|c|c|c|}
\hline LH  & x=0 & x=0.05 &  x=0.1 & x=0.15&x=0.2
 \tabularnewline \hline f=500& 87.22 & 79.23  & 72.33 & 66.44& 61.59
 \tabularnewline \hline f=1000& 27.84 & 25.12   & 22.55  & 20.11  & 17.83
 \tabularnewline \hline LRTH&M=0 & M=100 & M=300  & M=500& M=700
 \tabularnewline \hline f=500&0 & 1.9 & 12.5   & 22.4 & 29
 \tabularnewline \hline f=1000&0 & 0.10 & 0.86  & 2.07  &  3.36
\tabularnewline \hline
\end{tabular}\caption{For $m_\phi=400$ GeV, the cross section of the signal process at  $E_{cm}=14$
TeV for $f$ and $M$ in unit of GeV, cross sections in unit of $fb$.} \label{cross-section}
\par\end{centering}
\centering{}
\end{table}

From Table \ref{cross-section}, we can see that in the LH, when $m_S =400 GeV$, for the small
scale $f$, the the cross sections are larger than $49.7$ fb, the value for the $3\sigma$
confidence level. While for the LRTH, it  is dangerous
to reach the detectable level in the most parameter space.
In LH models, when $f$ is large, the production rates will be suppressed and smaller than  $49.7$ fb, which
will be hardly to probe. The cross sections, however, are also sensitive to the  parameters $s$ and
$s'$, this would give quite larger results if we fine tune the parameters.  When $s=s'=0.1$, for example,
the production can even arrive at $1000$fb.  However, this fine-tuning is not
what we want, since it only in a little parameter space and we should consider the confinements such as  Ref. \cite{litt-constr,0303236}.

In LH models, however, we can also consider the larger scalar mass, such as $600$ GeV, according to Table I and Table II in
Ref. \cite{11120086}, the cross sections before cuts is about $14$ fb, and the $S/\sqrt{B}$
is $8.77$ with the integral luminosity $300$ fb$^{-1}$. We calculate the rate of the
$\phi W$ production at $m_\phi=600$ GeV for $f=1000$ GeV and $s=0.1$, we just find that the
cross section can arive at about $9$ fb, which is close to $14$ fb.  So we can image that the signal
and backgrounds $S/\sqrt{B}$ should be large, at least larger than $3$ for such a large cross section for
 $m_\phi=600$ GeV. So we may conclude that
 for a larger scalar mass, the associated production could be more easily to be detected.

As for the other production modes of the charged higgs in LH and LRTH models, the pair production should be the most interesting one since the order may be large. For the two models,
the large SM backgrounds do not take too much luck to the detection,
just stated as Ref. \cite{lh-lrth-pheno}, it may only be possible for
the charged higgs to be produced in quite a narrow space.
The pair productions of the neutral higgs are also discussed \cite{lh-lrth-pheno}
and be also possible in a narrow parameter space.
Other production modes of the higgs in LH and LRTH models such as $ZH$,
$tH$ and $ZHH$ \cite{lh-lrth-pheno} are also studied.

If one wants to detect all these procedures list above, the common requirements are
that both the $f$ and the higgs masses are not too large, which are agreed in principle
with $W$ and the charged higgs associated production, which has been discussed in
this work. Larger $f$, e.g, $f>1000$ GeV, however, is preferred by current constraints, so
it may be an another interesting issue to consider a larger boson mass to carry out the detection of this signal.
As we have discussed, it may also be possible to be probed in a small parameter spaces.
Though $f$ is relatively large ($f>800$ GeV), we in this work show with a smaller $f$ ($f$ from $500$ to
$2000$ GeV), too, to see the impact of this parameter on the associated production.


\section{Conclusion and Summary}
We calculate the charged scalar production associated with a gauge boson $W$ in the LH models
and the LRTH realizations.
Comparing the two kinds of models, we can see that, at the LHC, the
$\phi W$ production in the LH models are larger than that in the LRTH models,
and the LH models should be more more possible to be detected at the LHC via the $\phi^+ W^-$ production.
From the discussion above, we can also conclude that, in LH models,
for a small $f$, in most parameter space of the LH model, the production rates can arrive at the
detectable level. But when $f$ is large, the suppression effect becomes strong, so it may difficult for LHC to
detect the signal. With a larger scalar mass, however, the signal will be a little easier to detect.


\section*{Acknowledgments} \hspace{5mm}
This work was supported by the National Natural Science
Foundation of China under the Grants No.11105125, No.11105124 and 11205023.

\end{document}